\documentclass[iop]{emulateapj}
\usepackage{graphicx}
\usepackage{blindtext}
\usepackage{amsmath}
\usepackage{mathtools}
\usepackage{multirow}
\usepackage{hyperref}
\hypersetup{
	colorlinks	= true,
	linkcolor	= red,
	urlcolor	= cyan,
	citecolor	= blue
}

\newcommand{\hst}{\mbox{HST}}
\newcommand{\wfc}{\mbox{WFC3}}
\newcommand{\stis}{\mbox{STIS}}
\newcommand{\grism}{\mbox{G141}}

\newcommand{\spitzer}{\mbox{Spitzer}}
\newcommand{\irac}{\mbox{IRAC}}

\newcommand{\taurex}{\mbox{$\mathcal{T}$-REx}}

\newcommand{\atlas}{\mbox{ATLAS}}
\newcommand{\phoenix}{\mbox{PHOENIX}}

\begin{document}

\title{A population study of gaseous exoplanets}

\author{A. Tsiaras\altaffilmark{1}, I.P. Waldmann\altaffilmark{1}, T. Zingales\altaffilmark{1,2}, M. Rocchetto\altaffilmark{1}, G. Morello\altaffilmark{1}, M. Damiano\altaffilmark{1,2}, K. Karpouzas\altaffilmark{3}, \\ G. Tinetti\altaffilmark{1}, L. K. McKemmish\altaffilmark{1}, J. Tennyson\altaffilmark{1}, S. N. Yurchenko\altaffilmark{1}}

\affil{$^1$Department of Physics \& Astronomy, University College London, Gower Street, WC1E6BT London, United Kingdom}
\affil{$^2$INAF--Osservatorio Astronomico di Palermo, Piazza del Parlamento 1, I-90134 Palermo, Italy}
\affil{$^3$Department of Physics, Section of Astrophysics, Astronomy and Mechanics, Aristotle University of Thessaloniki, 541 24 Thessaloniki, Greece}

\email{angelos.tsiaras.14@ucl.ac.uk}

\begin{abstract}
We present here the analysis of 30 gaseous extrasolar planets, with temperatures between 600 and 2400\,K and radii between 0.35 and 1.9\,$R_\mathrm{Jup}$. The quality of the \hst/\wfc\ spatially scanned data combined with our specialized analysis tools allow us to study the largest and most self-consistent sample of exoplanetary transmission spectra to date and examine the collective behavior of warm and hot gaseous planets rather than isolated case-studies. We define a new metric, the Atmospheric Detectability Index (ADI) to evaluate the statistical significance of an atmospheric detection and find statistically significant atmospheres around 16 planets out of the 30 analysed. For most of the Jupiters in our sample, we find the detectability of their atmospheres to be dependent on the planetary radius but not on the planetary mass. This indicates that planetary gravity plays a secondary role in the state of gaseous planetary atmospheres. We detect the presence of water vapour in all of the statistically detectable atmospheres, and we cannot rule out its presence in the atmospheres of the others. In addition, TiO and/or VO signatures are detected with 4\,$\sigma$ confidence in WASP-76\,b, and they are most likely present in WASP-121\,b. We find no correlation between expected signal-to-noise and atmospheric detectability for most targets. This has important implications for future large-scale surveys.
\end{abstract}

\keywords{methods: data analysis --- methods: statistical --- planets and satellites: atmospheres --- techniques: spectroscopic}

\maketitle

\section{INTRODUCTION} \label{sec:introduction}

We have progressed significantly from the first detections of atmospheric signatures in extrasolar planet atmospheres  \citep[e.g.][]{Charbonneau2002, Richardson2007, Tinetti20072007Natur.448..169T, Grillmair2008, Knutson2008, Redfield2008, Swain20082008Natur.452..329S} and are rapidly entering the era of comparative exoplanetology. Whilst individual case-studies of hot-Jupiters \citep[e.g.][]{Brogi2013, deKok2013, Deming2013, Konopacky2013, Mandell2013, Todorov2013, McCullough20142014ApJ...791...55M, Snellen2014, Stevenson20142014Sci...346..838S, Zellem20142014ApJ...796...48Z, Macintosh2015, Kreidberg2015, Iyer2016, Line20162016AJ....152..203L, Tsiaras20162016ApJ...832..202T} down to Neptune/Uranus \citep[e.g.][]{Stevenson2010, Fukui2013, Ehrenreich2014, Fraine2014, Knutson20142014Natur.505...66K, Morello20152015ApJ...802..117M} and super-Earths \citep[e.g.][]{Bean2010, Berta2012, Knutson20142014ApJ...794..155K, Kreidberg20142014Natur.505...69K, Demory20162016Natur.532..207D, Tsiaras20162016ApJ...820...99T} allow us to learn important properties of the planets analysed, we can only gain a limited insight into the global population and potential classifications of these foreign worlds. Population synthesis studies based on formation scenarios or statistics from the Kepler Space mission suggest a great diversity in the exoplanet population \citep[e.g.][]{Fortney2013, Lopez2014, Rogers2015, Parmentier2016}. To break current model degeneracies,  we need to access the chemical composition of these objects: this can be achieved through the observation of their atmospheres.

With the maturation of data analysis techniques for the Hubble/WFC3 camera (and other ground-based instruments), we are rapidly entering the stage of atmospheric surveys. A notable comparative study of 10 hot-Jupiters was presented last year \citep{Sing2016}.
For  large-scale studies to fulfill their promise of comparative planetology, two criteria must be met: 1) Homogeneity in data analysis: spectra need to be uniformly analyzed to mitigate biases; 2) Quantitative and homogeneous atmospheric modeling: quantitative analysis using atmospheric retrieval software applied to all spectra allows the exact statistical comparability between planetary and atmospheric parameters.

Here we present the analysis of 30 hot-Jupiters observed with the \hst/\wfc\ camera, in the spatially scanning mode, ranging from warm-Neptunes to very hot-Jupiters. Data were obtained from the publicly accessible pages of the NASA Mikulski Archive for Space Telescopes (MAST) archive. This paper contains the largest catalog of uniformly and quantitatively studied exoplanetary atmospheres to date, using the most precise observations currently available.

In the sections below, we present the data analysis and atmospheric retrieval frameworks used and discuss a new metric, the Atmospheric Detectability Index (ADI), for the quantitative assessment of the significance of the atmospheric signatures. We then use the ADI to search for potential correlations between the atmospheric features and basic planetary parameters, such as the size, temperature, mass etc. 

\section{DATA ANALYSIS} \label{sec:analysis}

We studied all the currently observed hot and gaseous planets with masses higher than 10\,M$_\Earth$ and estimated atmospheric absorption larger than three times the pre-calculated signal-to-noise ratio (pre-calculated $\mathrm{S/N} > 3$). The expected absorption at 1.4\,$\mu$m was calculated assuming an atmosphere with a  mean molecular weight of 2.3 amu and absorption features which sound five scale heights. The expected flux was calculated using the \wfc\ exposure time calculator. The planets included in the sample are: GJ\,436\,b, GJ\,3470\,b, HAT-P-1\,b, HAT-P-3\,b, HAT-P-11\,b, HAT-P-12\,b, HAT-P-17\,b, HAT-P-18\,b, HAT-P-26\,b, HAT-P-32\,b, HAT-P-38\,b, HAT-P-41\,b, HD\,149026\,b, HD\,189733\,b, HD\,209458\,b, WASP-12\,b, WASP-29\,b, WASP-31\,b, WASP-39\,b, WASP-43\,b, WASP-52\,b, WASP-63\,b, WASP-67\,b, WASP-69\,b, WASP-74\,b, WASP-76\,b, WASP-80\,b, WASP-101\,b, WASP-121\,b, and XO-1\,b. For some planets, other data sets using \hst/\stis, \spitzer/\irac\ and ground-based data exist \citep[e.g.][]{Danielski2014, Stevenson20142014AJ....147..161S, Snellen2014, Line20162016AJ....152..203L, Sing2016}. Here we restrict ourselves to \hst/\wfc\ data for reasons of comparability and homogeneity in the analysis. We note also that in the absence of any overlap in the wavelength ranges probed by \hst/\stis, \hst/\wfc \, and \spitzer/\irac\, an absolute calibration at the level of 10 to 100\,ppm between the different instruments is not guaranteed, making quantitative atmospheric retrievals sensitive to arbitrary offsets. 

Despite being eligible, we did not include in our sample some of the available transit observations as they were affected by different kinds of systematics. These observations were: a) the second transit of HAT-P-11\,b (ID: 12449, PI: D. Deming), due to the very large x-shifts of about 10 pixels, b) the first transit of HD\,149026\,b (ID: 14260, PI: D. Deming), as the spectrum was placed at the right edge of the detector, c) one transit of HAT-P-18\,b (ID: 14099, PI: T. Evans), due to a possible star spot occultation, d) two transits of XO-2\,b (ID: 13653, PI: C. Griffith), as the maximum flux per pixel exceeded the saturation level of 70,000 electrons, e) the third transit of GJ 3470\,b (ID: 13665, PI: B. Benneke), in which the spectrum was possibly contaminated close to the 1.4\,$\mu$m band.

From all the analyzed transit observations, the first \hst\ orbit was removed because of the strong systematics that affect it. Recently, \cite{Zhou2017} proposed a notable reduction method which also corrects for systematics in the first \hst\ orbit. A comparison between our pipeline and the approach proposed by \cite{Zhou2017} is beyond the scope of this paper, especially given that they have similar (nearly photon noise limited) performances. In some cases, a few spectroscopic images were also removed, as they were affected either by ``snowballs'' or by satellite trails. A complete list with the number of transit observations and \hst\ orbits used, as well as the references for the parameters used, can be found in Table \ref{tab:data}.

\begin{table}
	\small
	\center
	\caption{Proposal information for the data used in our analysis.}
	\label{tab:data}
	\setlength{\tabcolsep}{0.3em}
	\begin{tabular}{l | c c | c  c}
	\hline \hline
	Planet & Proposal & Proposal & {\footnotesize Transits} & {\footnotesize \hst\ } \\
	           &            ID &            PI &      {\footnotesize used} & {\footnotesize orbits} \\
	           &                &                 &                                       & {\footnotesize used} \\	           
	\hline
GJ\,436\,b & 11622 & Heather Knutson & 4 & 12 \\
GJ\,3470\,b & 13665 & Bjoern Benneke & 2 & 6 \\
HAT-P-1\,b & 12473 & David Sing & 1 & 4 \\
HAT-P-3\,b & 14260 & Drake Deming & 2 & 8 \\
HAT-P-11\,b & 12449 & Drake Deming & 1 & 3 \\
HAT-P-12\,b & 14260 & Drake Deming & 2 & 8 \\
HAT-P-17\,b & 12956 & Catherine Huitson & 1 & 4 \\
HAT-P-18\,b & 14260 & Drake Deming & 2 & 8 \\
HAT-P-26\,b & 14260 & Drake Deming & 2 & 8 \\
HAT-P-32\,b & 14260 & Drake Deming & 1 & 4 \\
HAT-P-38\,b & 14260 & Drake Deming & 2 & 8 \\
HAT-P-41\,b & 14767 & David Sing & 1 & 4 \\
HD\,149026\,b & 14260 & Drake Deming & 1 & 4 \\
HD\,189733\,b & 12881 & Peter McCullough & 1 & 6 \\
HD\,209458\,b & 12181 & Drake Deming & 1 & 4 \\
WASP-12\,b & 13467 & Jacob Bean & 3 & 12 \\
WASP-29\,b & 14260 & Drake Deming & 1 & 4 \\
WASP-31\,b & 12473 & David Sing & 1 & 4 \\
WASP-39\,b & 14260 & Drake Deming & 2 & 8 \\
WASP-43\,b & 13467 & Jacob Bean & 6 & 18 \\
WASP-52\,b & 14260 & Drake Deming & 1 & 3 \\
WASP-63\,b & 14642 & Kevin Stevenson & 1 & 7 \\
WASP-67\,b & 14260 & Drake Deming & 1 & 3 \\
WASP-69\,b & 14260 & Drake Deming & 1 & 3 \\
WASP-74\,b & 14767 & David Sing & 1 & 3 \\
WASP-76\,b & 14260 & Drake Deming & 1 & 4 \\
WASP-80\,b & 14260 & Drake Deming & 1 & 3 \\
WASP-101\,b & 14767 & David Sing & 1 & 4 \\
WASP-121\,b & 14468 & Thomas Evans & 1 & 4 \\
XO-1\,b & 12181 & Drake Deming & 1 & 4 \\
	\end{tabular}
\end{table}

\subsection{Reduction and calibration}

Our analysis started from the raw spatially scanned spectroscopic images, using our specialized software for the analysis of \wfc, spatially scanned spectroscopic images \citep{Tsiaras20162016ApJ...832..202T, Tsiaras20162016ApJ...820...99T}. The reduction process included the following steps: zero-read subtraction, reference pixels correction, nonlinearity correction, dark current subtraction, gain conversion, sky background subtraction, calibration, flat-field correction, and bad pixels/cosmic rays correction. In a broad sample like the current one, the possibility of observing additional sources in the field of view is high. Hence, we could not define the sky-area prior to the analysis, and the use of an automatic tool was necessary. The selected sky-area pixels were those with a flux level below a certain threshold -- twice the flux median absolute deviation (mad) from the median flux -- in all nondestructive reads. In cases where multiple transit observations were available (see  Table \ref{tab:data}), we calculated the position shifts by comparison with the first spectroscopic image of the first observation. This approach was followed to eliminate any systematic position shifts between the direct images of the different observations. While absolute calibration using the direct image has an uncertainty of $\pm$0.5\,pixels \citep{Kuntschner20092009wfc..rept...17K}, relative calibration can provide uncertainties of $\pm$0.005\,pixels \citep[Figure 14 in][]{Varley2017}.

\begin{table*}
	\small
	\center
	\caption{Parameters used in our analysis. The transit mid-time and depth are not reported as they are fitted in all cases as free parameters.\vspace{0.3cm}}
	\label{tab:parameters}
	\setlength{\tabcolsep}{0.4em}
	\begin{tabular}{l | c c c c | c c | c c c c c | c}
	\hline \hline
	Planet & [Fe/H]$_*$ & $T_*$ & $\log(g_*)$ & $R_*$         & $M_\mathrm{p}$     & $R_\mathrm{p}$     & $P$  & $i$  & $a/R_*$ & $e$ & $\omega$ & Reference \\
	           &                   & K         & cgs             & $R_\oplus$ & $M_\mathrm{Jup}$ & $R_\mathrm{Jup}$ & days & deg &              &         & deg          &                   \\ 
	\hline
GJ\,436\,b & 0.02 & 3416 & 4.843 & 0.455 & 0.08 & 0.366 & 2.64389803 & 86.858 & 14.54 & 0.1616 & 327.2 & \cite{Lanotte2014} \\
GJ\,3470\,b & 0.17 & 3652 & 4.78 & 0.48 & 0.043 & 0.346 & 3.3366487 & 88.88 & 13.94 & - & - & \cite{Biddle2014} \\
HAT-P-1\,b & 0.13 & 5975 & 4.45 & 1.15 & 0.53 & 1.36 & 4.46529 & 85.9 & 10.247 & - & - & \cite{Bakos2007} \\
HAT-P-3\,b & 0.27 & 5185 & 4.564 & 0.833 & 0.596 & 0.899 & 2.899703 & 87.24 & 10.59 & - & - & \cite{Torres2008} \\
HAT-P-11\,b & 0.31 & 4780 & 4.59 & 0.75 & 0.081 & 0.422 & 4.8878162 & 88.5 & 15.58 & 0.198 & 355.2 & \cite{Bakos2010} \\
HAT-P-12\,b & -0.29 & 4650 & 4.61 & 0.701 & 0.211 & 0.959 & 3.2130598 & 89 & 11.77 & - & - & \cite{Hartman2009} \\
HAT-P-17\,b & 0 & 5246 & 4.52 & 0.838 & 0.534 & 1.01 & 10.338523 & 89.2 & 22.63 & 0.342 & 201 & \cite{Howard20122012ApJ...749..134H} \\
HAT-P-18\,b & 0.1 & 4870 & 4.57 & 0.717 & 0.196 & 0.947 & 5.507978 & 88.79 & 16.67 & - & - & \cite{Esposito2014} \\
HAT-P-26\,b & -0.04 & 5079 & 4.56 & 0.788 & 0.057 & 0.549 & 4.234515 & 88.6 & 13.44 & - & - & \cite{Hartman20112011ApJ...728..138H} \\
HAT-P-32\,b & -0.04 & 6207 & 4.33 & 1.219 & 0.86 & 1.789 & 2.150008 & 88.9 & 6.05 & - & - & \cite{Hartman20112011ApJ...742...59H} \\
HAT-P-38\,b & 0.06 & 5330 & 4.45 & 0.923 & 0.267 & 0.825 & 4.640382 & 88.3 & 12.17 & - & - & \cite{Sato2012} \\
HAT-P-41\,b & 0.21 & 6390 & 4.14 & 1.683 & 0.8 & 1.685 & 2.694047 & 87.7 & 5.44 & - & - & \cite{Hartman2012} \\
HD\,149026\,b & 0.36 & 6160 & 4.278 & 1.368 & 0.359 & 0.654 & 2.87598 & 90 & 7.11 & - & - & \cite{Torres2008} \\
HD\,189733\,b & -0.03 & 5040 & 4.587 & 0.756 & 1.144 & 1.138 & 2.218573 & 85.58 & 8.81 & - & - & \cite{Torres2008} \\
HD\,209458\,b & 0 & 6065 & 4.361 & 1.155 & 0.685 & 1.359 & 3.524746 & 86.71 & 8.76 & - & - & \cite{Torres2008} \\
WASP-12\,b & 0.33 & 6360 & 4.157 & 1.657 & 1.47 & 1.9 & 1.0914203 & 83.37 & 3.039 & - & - & \cite{Collins2017} \\
WASP-29\,b & 0.11 & 4800 & 4.54 & 0.808 & 0.244 & 0.792 & 3.922727 & 88.8 & 12.415 & - & - & \cite{Hellier2010} \\
WASP-31\,b & -0.2 & 6302 & 4.308 & 1.252 & 0.478 & 1.549 & 3.4059096 & 84.41 & 8 & - & - & \cite{Anderson2011} \\
WASP-39\,b & -0.12 & 5400 & 4.503 & 0.895 & 0.28 & 1.27 & 4.055259 & 87.83 & 11.647 & - & - & \cite{Faedi2011} \\
WASP-43\,b & -0.05 & 4400 & 4.65 & 0.67 & 1.78 & 0.93 & 0.813475 & 82.6 & 5.124 & - & - & \cite{Hellier2011} \\
WASP-52\,b & 0.03 & 5000 & 4.582 & 0.79 & 0.46 & 1.27 & 1.7497798 & 85.35 & 7.401 & - & - & \cite{Hebrard2013} \\
WASP-63\,b & 0.08 & 5550 & 4.01 & 1.88 & 0.38 & 1.43 & 4.37809 & 87.8 & 6.773 & - & - & \cite{Hellier2012} \\
WASP-67\,b & -0.07 & 5200 & 4.5 & 0.87 & 0.42 & 1.4 & 4.61442 & 85.8 & 12.835 & - & - & \cite{Hellier2012} \\
WASP-69\,b & 0.144 & 4715 & 4.535 & 0.813 & 0.26 & 1.057 & 3.8681382 & 86.71 & 11.953 & - & - & \cite{Anderson2014} \\
WASP-74\,b & 0.39 & 5970 & 4.18 & 1.64 & 0.95 & 1.56 & 2.13775 & 79.81 & 4.861 & - & - & \cite{Hellier2015} \\
WASP-76\,b & 0.23 & 6250 & 4.128 & 1.73 & 0.92 & 1.83 & 1.809886 & 88 & 4.012 & - & - & \cite{West2016} \\
WASP-80\,b & -0.13 & 4143 & 4.663 & 0.586 & 0.538 & 0.999 & 3.06785234 & 89.02 & 12.63 & - & - & \cite{Triaud2015} \\
WASP-101\,b & 0.2 & 6380 & 4.345 & 1.29 & 0.5 & 1.41 & 3.585722 & 85 & 8.445 & - & - & \cite{Hellier2014} \\
WASP-121\,b & 0.13 & 6460 & 4.242 & 1.458 & 1.183 & 1.865 & 1.2749255 & 87.6 & 3.754 & - & - & \cite{Delrez2016} \\
XO-1\,b & 0.02 & 5750 & 4.509 & 0.934 & 0.918 & 1.206 & 3.941534 & 88.81 & 11.55 & - & - & \cite{Torres2008} \\
	\end{tabular}
\vspace{0.4cm}
\end{table*}

\paragraph{HD\,189733\,b} During the spatial scans of HD\,189733\,b the spectrum was shifted above the upper edge of the detector. Hence only the first three nondestructive reads were used from the forward scans and only the last five from the reverse scans. Due to the different exposure times, forward and reverse scans were processed independently as two different transit observations.

\subsection{Light-curves extraction} 

Following the reduction process, we extracted the flux from the spatially scanned spectroscopic images to create the final transit light-curves per wavelength band. We considered one broad band (white) covering the whole wavelength range in which the G141 grism is sensitive (1.088 -- 1.68\,$\mu$m), and two different sets of narrow bands (spectral). The resolving power of each set of narrow bands at 1.4\,$\mu$m was 50 (low) and 70 (high), respectively. In both sets, the widths of the narrow bands were varying between 0.0188 and 0.0390\,$\mu$m, in a way that the flux of a sun-like star would be equal in all the bands. The choice of the narrow bands sizes ensured an approximately uniform S/N across the planetary spectrum. We extracted our final light-curves from the differential nondestructive reads, a commonly used technique \citep{Deming2013}. In this way we also avoid any potential overlap of different spectra in cases where close companions exist.

\subsection{Limb darkening coefficients}

\begin{figure*}
	\centering
	\includegraphics[width=\textwidth]{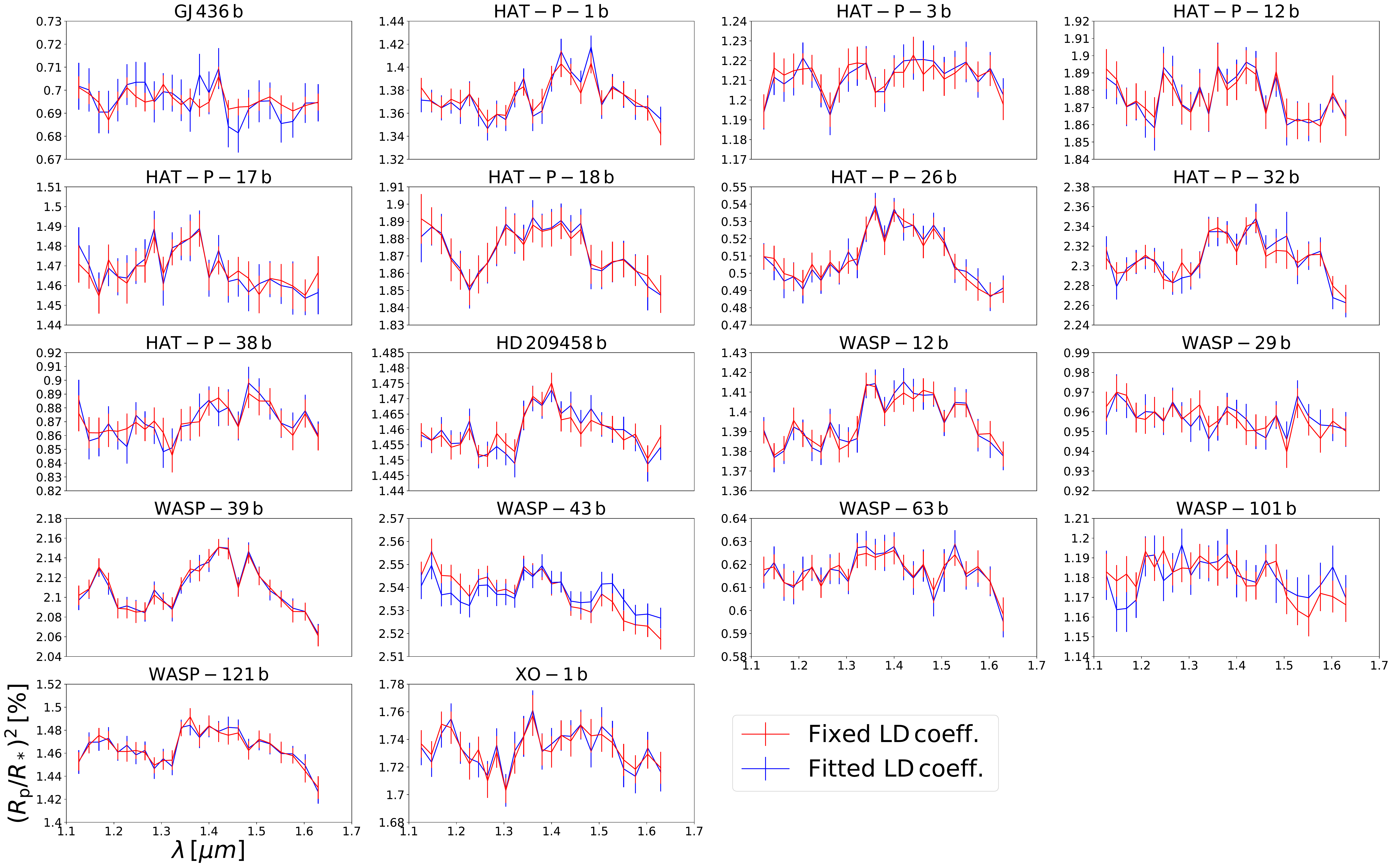}
	\caption{Comparison between the spectra extracted using fixed limb-darkening coefficients (red) and a fitted linear limb-darkening coefficient (blue). The only noticeable difference is for the case of WASP-43\,b.}
	\label{fig:comp_ld}
\end{figure*}

We modeled the stellar limb darkening effect using the nonlinear formula proposed by \cite{Claret2000}. The coefficients were fitted on the specific intensity profiles, evaluated at 100 angles, directly computed from the \atlas\ model \citep{Howarth2011}, for stars with effective temperatures higher than 4000\,K, or \phoenix\ \citep{Allard2012} model, for stars with effective temperature lower than 4000\,K, convoluted with the throughput of the \grism\ grism of the \wfc\ camera. The stellar parameters used can be found in Table \ref{tab:parameters}.

Fitting the limb darkening coefficients directly to the light curves (together with the other transit and instrumental parameters) is not an option, because of the many parameter degeneracies. This limitation applies particularly valid to \hst\ observations, as they present periodic gaps during the transit events.

A detailed study by \cite{Morello2017} shows that uncertainties in the stellar models do not significantly affect the atmospheric spectra in the \wfc\ passband. For a subset of planets, where fitting a linear limb-darkening coefficient was possible, we tested this option and found that it is not affecting the shape of the final spectrum but may introduce only a vertical offset. The only exception was WASP-43\,b (see Figure \ref{fig:comp_ld}).
 
\subsection{White light-curves fitting} \label{white_fitting}

As in previous observations with \wfc\ \citep[e.g.][]{Kreidberg2015, Line20162016AJ....152..203L, Evans2016, Wakeford20172017ApJ...835L..12W} our extracted raw white light-curves were affected by two kinds of time-dependent systematics, the long-term and short-term ``ramps''. The first is affecting each \hst\ visit and has a linear behavior, while the second affects each orbit and has an exponential behavior. Additional systematics that cannot be described by the above functional forms are also very common \citep{Wakeford2016}. To account for these effects we fitted a model for the systematics simultaneously with the transit model \citep{Kreidberg20142014Natur.505...69K, Tsiaras20162016ApJ...832..202T}. We varied the parameters of the short-term ramp for the first orbit in the analyzed time-series, as in many cases the first orbit was  affected in a different way compared to the other orbits. In addition, the parameters of the exponential short-term ramp also varied for the mid-orbit ramps caused by buffer dumps during an \hst\ orbit. Finally, forward and reverse scans were combined together by using a different normalization factor to account for the shift between them.

After an initial fit we scaled-up the uncertainties on the individual data points, in order for their median to match the standard deviation of the residuals, and fitted again. In this way we adopted more conservative values for the uncertainties of the fitted parameters, taking into account the systematics that were not described by our functional form.
 
\begin{figure}
	\centering
	\includegraphics[width=\columnwidth]{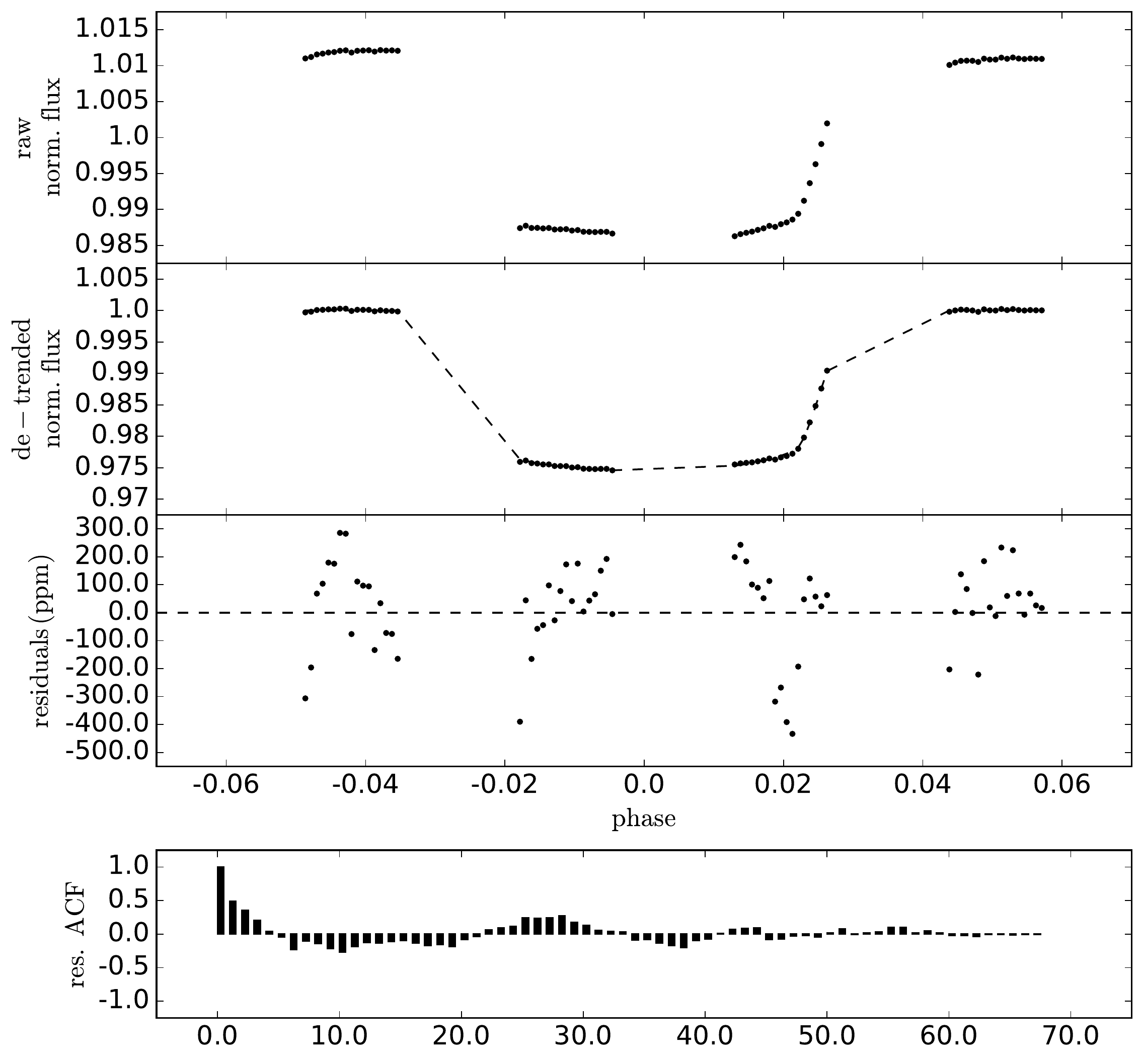}
	\caption{Results from the analysis of the white light-curve of HAT-P-32\,b. Top panel: Normalized raw light-curve. Second panel: Light-curve divided by the best-fit model for the systematics. Third panel: Fitting residuals. Bottom panel: Autocorrelation function of the residuals. This planet belongs to this group which is affected by additional systematics (group c in Section \ref{white_fitting}) and as we can see in the lower two panels the residuals are not following a gaussian distribution.}
	\label{fig:hatp32}
\end{figure}

All the white light-curves were fitted for the $R_\mathrm{p}/R_*$ and $T_0$ parameters, using fixed values for the $P$, $e$ and $\omega$ parameters, as reported in the literature (see Table \ref{tab:parameters}). Concerning the $i$ and $a/R_*$ parameters, the planets in our sample can be divided in three categories: 

a) successfully fitted\footnote{we consider a fit to be successful if the autocorrelation of their white light-curve residuals is below 0.3 (Figure \ref{fig:residuals}).} with literature values for $i$ and $a/R_*$: this category includes the majority of the white light-curves (GJ\,3470\,b, HAT-P-11\,b, HAT-P-26\,b, HAT-P-38\,b, HAT-P-41\,b, HD\,149026\,b, WASP-29\,b, WASP-31\,b, WASP-43\,b, WASP-67\,b, WASP-69\,b, WASP-74\,b, WASP-80\,b, WASP-101\,b).

b) successfully fitted with $i$ and $a/R_*$ as free parameters: this category includes those light-curves that showed additional systematics when the literature values for $i$ and $a/R_*$ were used, but corrected by fitting for $i$ and $a/R_*$ (HAT-P-1\,b, HAT-P-3\,b, HAT-P-12\,b, HAT-P-18\,b, WASP-39\,b, and XO-1\,b).

c) other effects: this category includes those light-curves that showed additional systematics when the literature values for $i$ and $a/R_*$ were used, but could not be corrected by fitting for $i$ and $a/R_*$. For these planets we finally decided to adopt the literature values for $i$ and $a/R_*$ if either the transit ingress or egress was not observed (HAT-P-32\,b shown in Figure \ref{fig:hatp32}, HD\,189733\,b, HD\,209458\,b, WASP-12\,b, WASP-52\,b, WASP-76\,b and WASP-121\,b) or the fitted values for $i$ and $a/R_*$ if both the transit ingress and egress were observed (GJ\,436\,b, HAT-P-17\,b, WASP-63\,b).

The higher systematic residuals in the third category of light curves (above 1.5 times the expected photon-noise limited residuals, Figure \ref{fig:residuals}) could be due to non-optimal sets of stellar limb darkening coefficients. The most likely causes of discrepancy between predicted and observed limb darkening coefficients are, for the cooler stars, stellar activity \citep[e.g.][]{Csizmadia2013} or inaccurate chemical models \citep[e.g.][]{Allard2012}, and, for the hotter stars, the use of a plane-parallel approximation rather than full spherical geometry \citep{Hayek2012, Morello2017}. We tested two different approaches to reduce the systematic noise in the residuals by changing the limb darkening coefficients, similar to those suggested by \cite{Howarth2017}: 1) fitting for a linear limb-darkening coefficient; 2) calculating the coefficients from stellar models with different temperatures. In this way, the resulting transit depths may vary by $\sim$\,100\,ppm (2--3\,$\sigma$).

\subsection{Spectral light-curves fitting} \label{sec:spectral_fit}

Finally, we fitted the spectral light-curves using the divide-white method introduced by \cite{Kreidberg20142014Natur.505...69K}, where the white light-curve was used as a comparison source, with the addition of a normalization factor and a wavelength-dependent slope, linear with time. In the same way as for the white light-curves, we performed an initial fit and then scaled-up the uncertainties on the individual data points based on the standard deviation of the residuals, and fitted again. The $i$ and $a/R_*$ parameters are fixed to the literature values or to the best-fit values obtained for the relevant white light curves. Concerning the fitting of the spectral light-curves, the wavelength-dependent slope was not correlated with the $R_\mathrm{p}/R_*$ parameter, despite the strength of the slope. The only exception was HAT-P-17\,b, as no observations after the transit were included in this data set. However, the strength of the slope was insignificant throughout the spectrum of HAT-P-17\,b ($< 1\, \sigma$). For each planet, two final spectra were extracted at different resolutions (high and low, from the two sets of narrow bands). For the cases were multiple transit observations were available, the final spectra were the weighted average of the individual spectra, corrected for potential offsets in the white light-curve depth from one transit to another.

\begin{figure}
	\centering
	\includegraphics[width=\columnwidth]{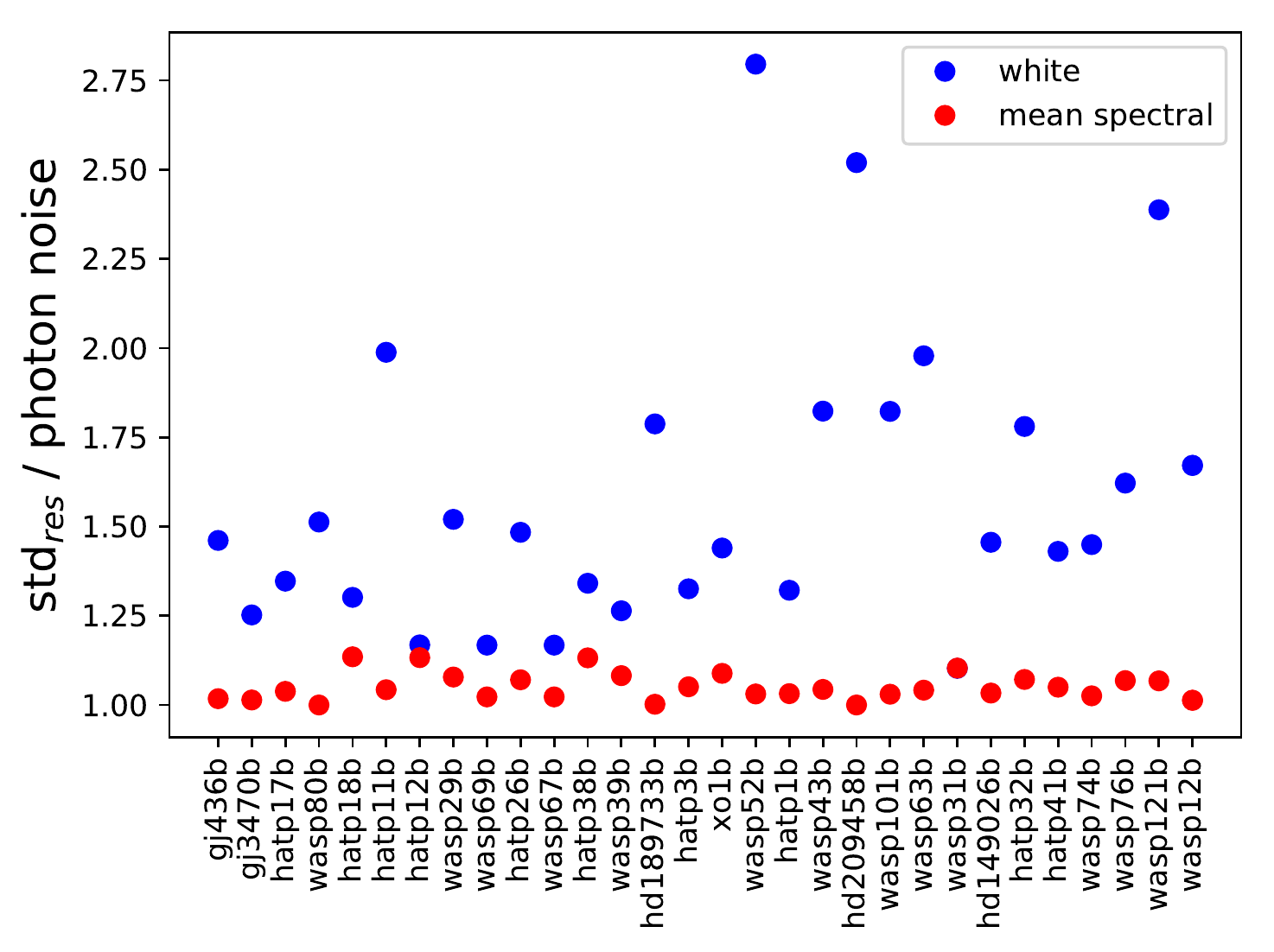}
   	\includegraphics[width=\columnwidth]{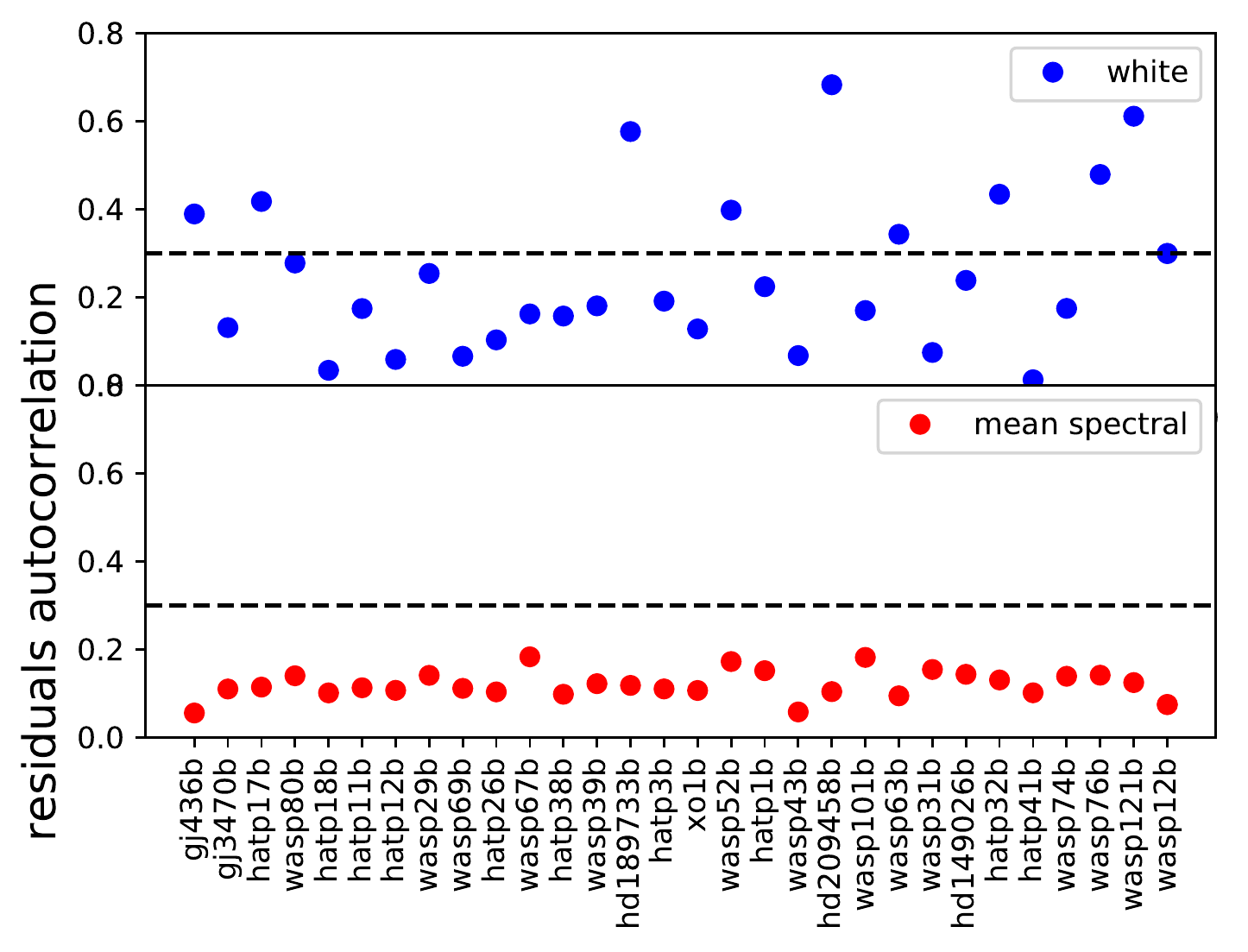}
	\caption{Standard deviation (top) and autocorrelation (bottom) of the fitting. These metrics indicate that while the white light-curves (blue) are subjected to remaining signals in their residuals, the divide-white method used here is efficient in removing those signals from the spectral light-curves (red).}
	\label{fig:residuals}
\end{figure}

In Figure \ref{fig:residuals} we can see that for the spectral light-curves, the standard deviation of the residuals is on average 1.05 (1.17 in the worst case) times the the expected photon-noise. In addition, the residuals autocorrelation is below 0.2. These metrics indicate that while the white light-curves are subjected to remaining signals in their residuals, the divide-white method used here is efficient in removing those signals from the spectral light-curves. Also, the tests with different limb-darkening coefficients show that the effect of the signals not fitted to the white light-curves can cause arbitrary offsets in the final spectra, but not change their shape. Hence the uncertainties reported for the spectra are referring to the relative transit depths, not including the uncertainties on the white light-curve depth.

The only exception was WASP-43\,b, for which different spectral slopes were obtained with different sets of limb darkening coefficients. For this planet, our original spectrum obtained showed a decreasing trend towards longer wavelengths. We found that, when fitting for a linear limb-darkening coefficient, the trend in the spectrum is less strong (Figure \ref{fig:comp_ld}) and in agreement with the literature \citep{Kreidberg20142014ApJ...793L..27K}. We report this as the final spectrum of WASP-43\,b.

\paragraph{WASP-80\,b}

From the spectrum of  WASP-80\,b, one data point at 1.4\,$\mu$m was excluded as it was contaminated by the zeroth order of the spectrum of a nearby source.

\paragraph{WASP-12\,b}

The spectrum of WASP-12\,b was contaminated by a very close companion. To correct for this effect, we used the starring-mode spectroscopic images included in the data set. From those images, we calculated a dilution factor, which we then used to correct the spectra \citep{Kreidberg2015}.


\section{ATMOSPHERIC MODELING} \label{sec:modeling}

The observed spectra were fitted using the Bayesian atmospheric retrieval framework \taurex\ \citep{ Waldmann20152015ApJ...813...13W, Waldmann20152015ApJ...802..107W}. \taurex\ fully maps the atmospheric correlated parameters retrieved from the observed spectra through the use of nested sampling \citep{Skilling2006, Feroz2009}. We modeled the transmission spectra using a variety of possible molecular opacities, namely H$_2$O, CH$_4$, CO, CO$_2$, NH$_3$, TiO, and VO. For most planets, water vapour is the only detectable signal together with clouds/hazes. However, TiO and VO were detected in WASP-76\,b with a 4.0\,$\sigma$ significance and are suggestive (but not significantly detected) in WASP-121\,b. Below, we  briefly describe the priors adopted, the general atmospheric parameterizations, opacity sources and cloud parameterization. All input parameters and full model outputs for each planet can be found in the data pack accompanying this paper. 

\subsection{General setup} 

The atmospheres of the planets analyzed here were simulated to range from 10$^{-4}$ to 10$^6$\,Pa and sampled uniformly in log-space by 100 atmospheric layers. We tested for potential under-sampling of the atmosphere by running test retrievals at 250 and 50 layers. No significant degradation of retrieval accuracy for \hst/\wfc\ data could be found. Each trace-gas abundance  was allowed to vary from 10$^{-8}$ to 10$^{-1}$ in volume mixing ratios (log-uniform prior) for hot-Jupiters  and 10$^{-8}$ to 1.0 for Neptunes (i.e. HAT-P-11\,b). From here forth, all priors are assumed to be uniform unless specified otherwise. We calculated planetary equilibrium temperatures  $T_p$ assuming geometric albedos varying from 0.6 to zero and emissivity from 0.5 to 1 to calculate the temperature prior range (as shown in equation \ref{eq:equ_temp}): 

\begin{equation}
T_p = T_*\left( \frac{R_*}{2a} \right)^{1/2} \left( \frac{1 - A}{ \varepsilon} \right)^{1/4}
\label{eq:equ_temp}
\end{equation}

where $R_*$ is the stellar radius, $a$ is the semi-major axis, $A$ is the geometric albedo and $\varepsilon$ is the planetary emissivity. For our temperature priors we used the $\left[ T_p(A=0.6, \varepsilon=1) -500\,K, \ T_p(A=0, \varepsilon=0.5) +500\,K \right]$ range. We adopted a wide temperature prior to allow for significantly cooler terminator temperatures compared to the expected equilibrium temperatures. Due to the short wavelength coverage of the HST/WFC3 instrument, we typically only probe a very restricted range of the planet's temperature-pressure profile.

An isothermal temperature-pressure profile was assumed. While this is an oversimplification and can lead to retrieval biases \citep{Rocchetto2016}, the restrictive wavelength range of 1.1 to 1.8\,$\mu$m does not allow the differentiation of an isothermal from a more complex profile. We adopted the planetary radius uncertainties reported in the literature as prior bounds and corrected them if needed. 

\subsection{Opacity sources}

Initially, exploratory retrievals were run to include a wide range of molecular opacities: H$_2$O, HCN, NH$_3$, CH$_4$, CO$_2$, CO, NO, SiO, TiO, VO, H$_2$S, and C$_2$H$_2$. No significant contributions were found but for H$_2$O, TiO and VO. We hence restricted further retrievals to a smaller set of molecules: H$_2$O \citep{Barber2006}, CO \citep{Rothman2010}, CO$_2$ \citep{Rothman2010}, CH$_4$ \citep{Yurchenko2014} and NH$_3$ \citep{Yurchenko2011}. VO \citep{Mckemmish2016} and TiO (McKemmish in prep.) were added to the mix for planets with equilibrium temperatures exceeding 1400\,K. Tau-REx is designed to operate with either absorption cross-sections or correlated-k coefficients. Both cross-sections and k-tables were computed from very high-resolution ($\mathrm{R}>10^6$) cross-sections, which in turn were calculated from molecular line lists obtained from ExoMol \citep{Tennyson2016}, HITEMP \citep{Rothman2010} and HITRAN \citep{Rothman2013}. Temperature and pressure dependent line-broadening was included, taking into account J-dependence where available \citep{Pine1992}. The absorption cross-sections were then binned to a constant resolution of $\mathrm{R}=15000$ and the transmission forward models were calculated at this resolution before binning to the resolution of the data. Given the resolutions, wavelength range and uncertainties of the data at hand, we find no differences between the use of cross-section and k-tables in the final retrieval results. Rayleigh scattering and collision induced absorption of H$_2$-H$_2$ and H$_2$-He was also included \citep{Borysow2001, Borysow2002, Rothman2013}.

\subsection{Cloud parameterization}

A variety of cloud parameterizations of varying complexity exist in the context of atmospheric retrieval studies \citep[e.g.][]{Benneke2012, Line20162016ApJ...820...78L, Barstow2013, Griffith2014}. Here we adopted the parameterization of \cite{Lee2013}, which also finds implementation in an atmospheric retrieval context in \cite{Lavie2017}. In transmission spectroscopy, the cloud optical depth as function of wavelength, $\tau_{c1,\lambda}$, is given by:
\begin{equation}
\tau_{c1,\lambda} = \int _0 ^{l(z)} Q_{ext,\lambda} \pi \alpha ^2 \chi_c (z') \rho_N (z') dl
\end{equation}

\noindent where z is the height in the atmosphere, $\alpha$ is the particle size of the cloud/haze, $dl$ is the path length through the atmosphere, $\chi_c$ is the cloud mixing-ratio, $\rho_N$ is the atmospheric number density, and $Q_{ext,\lambda}$ is the cloud extinction coefficient given by:
\begin{equation}
Q_{ext,\lambda} = \frac{5}{Q_0 x^{-4} + x^{0.2}}
\end{equation}

\noindent where $x=2 \pi \alpha / \lambda$ and $Q_0$ determines the peak of $Q_{ext,\lambda}$. This can be understood as a cloud compositional parameter \citep{Lee2013}. For $\alpha \ll \lambda$, the formalism reduces to pure Rayleigh scattering. In addition to the above, we implemented an optically thick grey-cloud cover, parameterized as follows:
\begin{equation}
\tau_{c2} = {\begin{cases}1, \ \mathrm{if} \ P < P_\mathrm{cloud-top}  \\ 0, \ \mathrm{otherwise} \end{cases}}
\end{equation}

\noindent where $P_\mathrm{cloud-top}$ is the cloud-top pressure. This dual parameterization allowed us to model optically thick cloud decks with a semi-transparent, hazy, atmosphere above $P_\mathrm{cloud-top}$. 

We initially kept $Q_0$, $\chi_c$, $\alpha$ (called R$_{cloud}$ in our retrieval corner plots), and $P_\mathrm{cloud-top}$ as free cloud parameters but found \hst/\wfc\ data to be insufficient to constrain $Q_0$. In initial tests, we varied $Q_0$ with a prior range from 0 - 100 but found $Q_0$ to be unconstrained by the data. We have therefore fixed $Q_0$ to the median value of 50 henceforth. We have found that uncertainty induced by either varying or fixing $Q_0$ is negligible given the quality of the data at hand. We set a log-uniform prior of $\chi_c$ ranging from 10$^{-40}$ to 10$^{-10}$, particle size from 10$^{-5}$ to 10\,$\mu$m and cloud top-pressure from 10$^{-4}$ to 10$^6$\,Pa \citep{Lee2013}.

\subsection{Free parameters and model selection} 

In the end, we had 10-12 free parameters: five molecular abundances (seven when TiO \& VO were included), temperature, planet radius and three cloud-deck parameters. Each one of the two spectra per planet at different resolutions was retrieved, yielding 60 retrievals in total. However, we found no difference between the information retrieved from the two spectra at different resolution. The results reported are from the low resolution spectra.

\subsection{Atmospheric Detectability Index (ADI)}

In order to quantify the detection significance of an atmosphere, we devised the Atmospheric Detectability Index (ADI). The ADI is the positively defined Bayes Factor between the nominal atmospheric model ($M_\mathrm{N}$) and a flat-line model ($M_\mathrm{F}$). As stated above, the nominal model contains molecular opacities, cloud/haze opacities ($\tau_{c1,\lambda}, \tau_{c2}$) collision induced absorption of H$_2$-H$_2$/H$_2$-He and Rayleigh scattering. Other free parameters are the planet radius, $R_\mathrm{p}$, and the temperature of the isothermal TP-profile, $T_\mathrm{iso}$. The flat-line model contains only grey-cloud opacities, $\tau_{c2}$, $R_\mathrm{p}$ and $T_\mathrm{iso}$. This parameterization always results in a flat-line spectrum but includes the model degeneracies found between cloud top-pressure, planet-radius and temperature. This way we capture both cloudy as well as clear sky scenarios. As the ADI is a fully Bayesian model selection metric, we naturally impose Occam's razor to our atmosphere detection significance. 

We obtained the Bayesian evidence of our nominal model, $E_\mathrm{N}$, and of the pure-cloud/no-atmosphere model, $E_\mathrm{F}$, and calculated the ADI as follows:
\begin{equation}
ADI = {\begin{cases} \log (E_\mathrm{N}) - \log (E_\mathrm{F}) , \ \mathrm{if} \ \log (E_\mathrm{N}) > \log (E_\mathrm{F})  \\ 0, \ \mathrm{otherwise} \end{cases}}
\end{equation}

The ADI is a positively defined metric and equivalent to the logarithmic Bayes Factor \citep{Kass1995} where $\log (E_\mathrm{N}) > \log (E_\mathrm{F})$.

\section{RESULTS AND DISCUSSION} \label{sec:results}

\subsection{Atmospheric detectability}

\begin{figure*}
	\centering
	\includegraphics[width=0.9\textwidth]{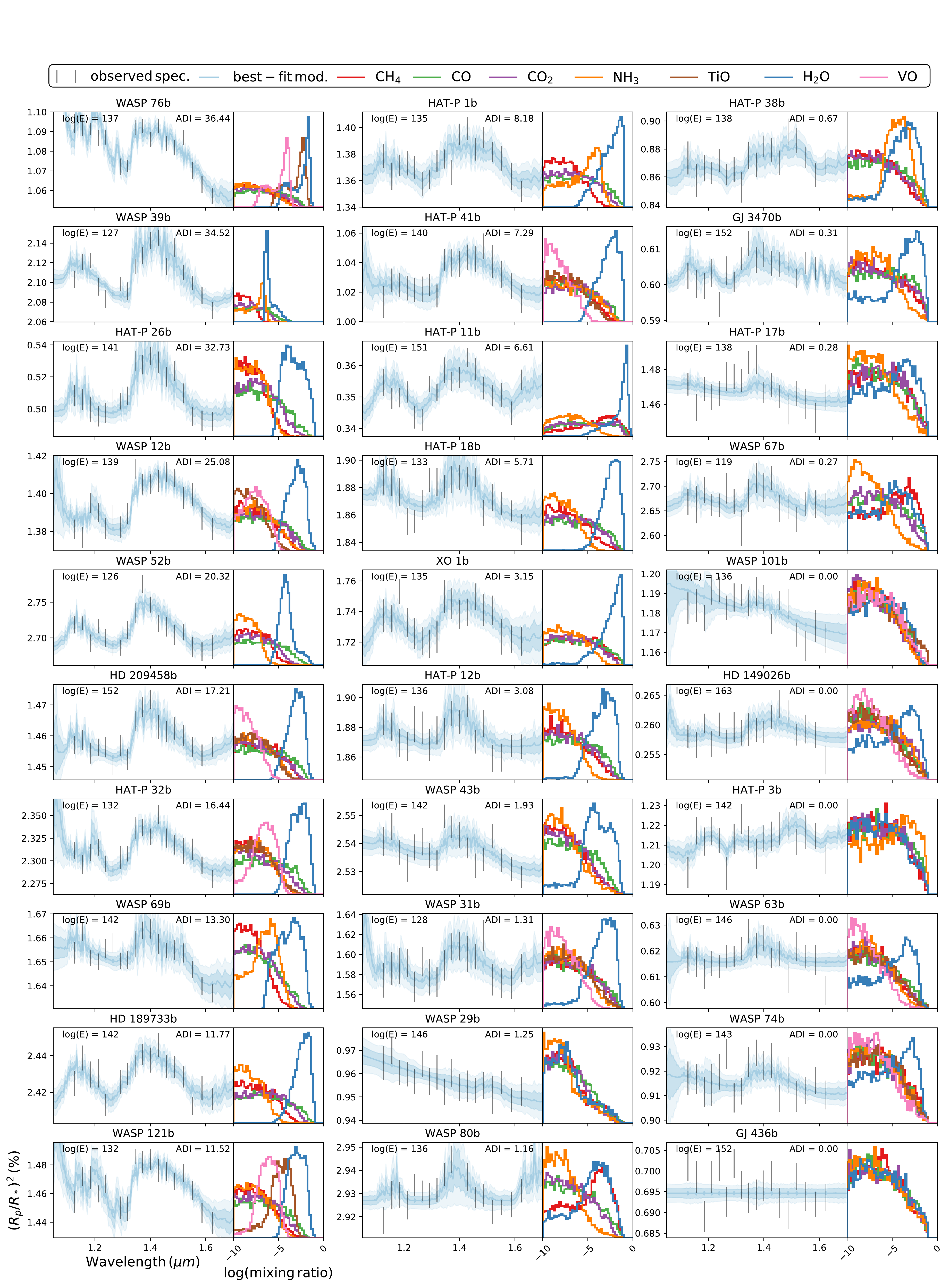}
	\caption{Atmospheric modeling results for all 30 planets in the sample. The planets are ordered based on the ADI index. The Bayesian evidence, log(E), of the best-fit model for each planet is also reported. Each panel shows, at left, the spectrum and the best-fit model and, at right, the posterior distributions of the abundances of the different molecules fitted.}
	\label{fig:all_results}
\end{figure*}

The low-resolution spectra obtained for all the planets in our sample are included in the on-line database (see following section). The ADI index has been reported for all the planets in Figure \ref{fig:all_results} and Table \ref{tab:all_results}. The spectra in Figure \ref{fig:all_results} are ordered by decreasing ADI. 

Given the definition of the ADI index in the previous section, an atmosphere is detected at 3\,$\sigma$ and 5\,$\sigma$ level for ADIs above 3 and 11 respectively. In our sample we find that 16 out of 30 planets feature statistically significant atmospheres, with ADIs higher than 3. While parameter constraints of atmospheric models for many of the planets with ADIs lower than 3 can be significant, indicating the presence of water (WASP-80\,b, WASP-43\,b, HAT-P-12\,b, HAT-P-38\,b, WASP-31\,b, WASP-63\,b, GJ\,3470\,b, WASP-67\,b, WASP-74\,b), the model as a whole is not. Hence, ADIs below 3 signify atmospheric nondetections, as the spectral feature amplitudes are insufficient (given the uncertainties in the data) to favor the more complex atmospheric model, $M_\mathrm{N}$ over the lower dimensional flat-line model $M_\mathrm{F}$. To verify the presence of water in these planets, additional observations are necessary. We have to note here that for WASP-43\,b the presence of water has been confirmed using additional observations during the eclipse of the planet \citep{Kreidberg20142014ApJ...793L..27K}. By adopting the ADI, we were able to draw several important conclusions about this population of exoplanets and spectroscopic observations of exoplanets in general.

\begin{figure*}
	\centering
	\includegraphics[width=0.6\textwidth]{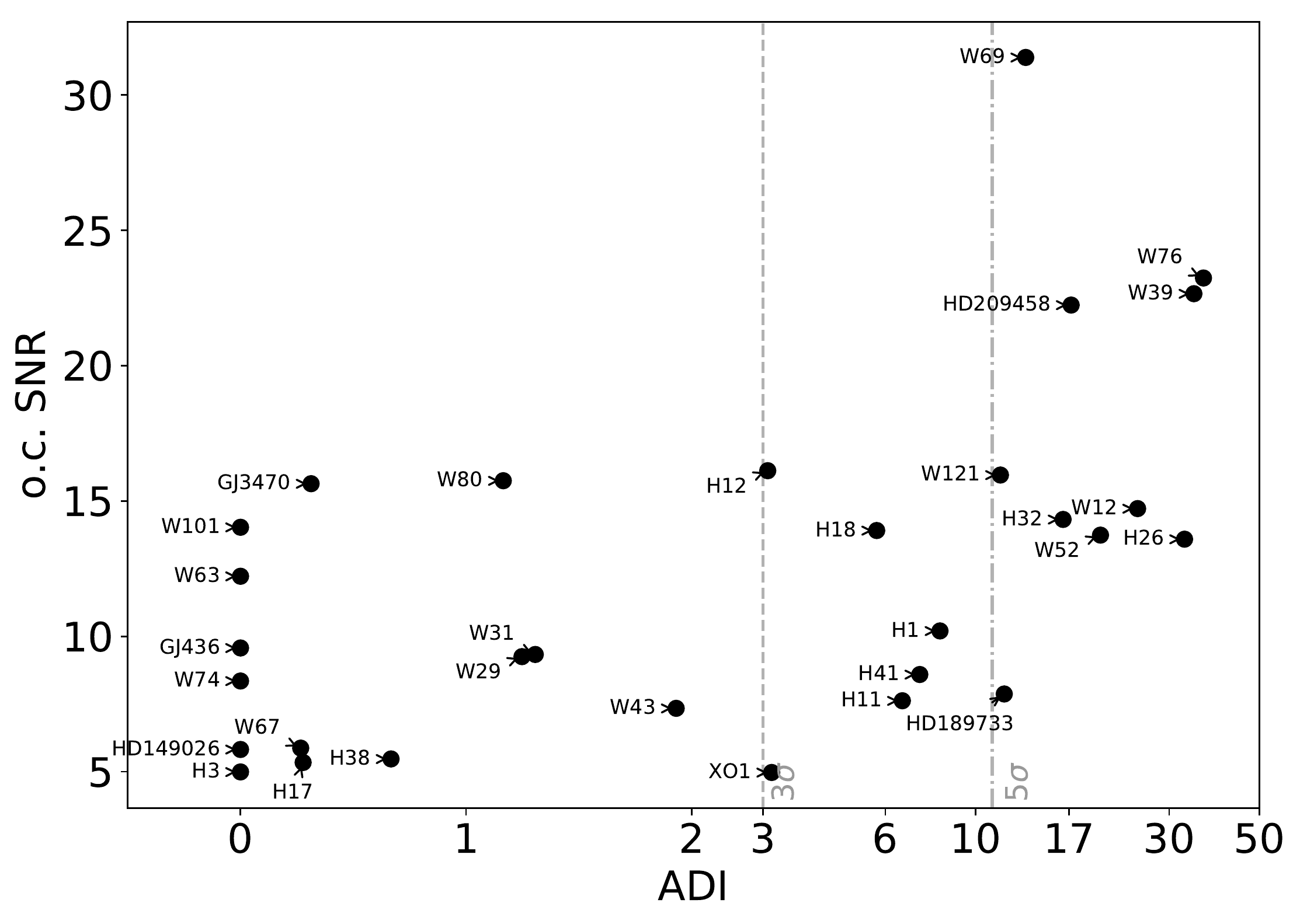}
	\caption{The o.c. S/N as a function of the ADI shows that planets with o.c. $\mathrm{S/N} > 20$ are always detectable but no correlation between ADI and o.c. S/N can be found for planets with o.c $\mathrm{S/N} < 20$.}
	\label{fig:adi_snr}
\end{figure*}

Previous population studies suggested that the observed spectra do not show the strong molecular features expected for a clear sky atmosphere \citep{Iyer2016, Sing2016}. Interestingly, even in this larger sample with all the planets expected to feature some modulations given the precision of the observations, the ADI does not correlate with the pre-calculated S/N. To exclude any observational biases we repeated the S/N calculation using the median uncertainty of our final observed low-resolution spectra instead of the pre-calculated uncertainties, we will refer to this quantity as observationally-corrected S/N (o.c. S/N). Interestingly, we find that for the planets with an o.c. S/N below 20, the ADI index is not correlated to the o.c. S/N (Figure \ref{fig:adi_snr}). In this regime we can find planets that scored highly on paper in terms of potential detections of atmospheric features but turned out to be difficult to interpret (e.g. WASP-101\,b), and planets that appeared relatively challenging to observe on paper but delivered very solid detections (e.g. HAT-P-11\,b). This absence of predictability showcases the need for exploratory observations prior to major time investments with large-scale facilities such as the JWST.

\begin{table*}
	\centering
	\small
	\center
	\caption{Observationally-corrected S/N, ADI, and main retrieval results (maximum a-posterior).}
	\label{tab:all_results}
	\setlength{\tabcolsep}{0.4em}
	\begin{tabular}{l | c c c c c c}
	\hline \hline
	Planet & o.c. S/N & ADI & $R_\mathrm{p}$ & $T_\mathrm{p}$ & $\log_{10} (P_\mathrm{cloup})$ & $\log_{10}$(H$_2$O) \\
	& & & $R_\mathrm{Jup}$ & K & Pa &  \\ 
	\hline
GJ\,436\,b & 9.57 & 0.00 & 0.37 $\pm$ 0.01 & 238.25 $\pm$ 188.69 & 1.22 $\pm$ 2.12 & --6.74 $\pm$ 2.70 \\ 
GJ\,3470\,b & 15.64 & 0.31 & 0.36 $\pm$ 0.01 & 243.68 $\pm$ 135.42 & 2.12 $\pm$ 1.57 & --4.87 $\pm$ 2.91 \\ 
HAT-P-1\,b & 10.20 & 8.18 & 1.29 $\pm$ 0.03 & 1017.09 $\pm$ 386.57 & 3.33 $\pm$ 1.35 & --2.68 $\pm$ 1.22 \\ 
HAT-P-3\,b & 4.99 & 0.00 & 0.89 $\pm$ 0.01 & 843.00 $\pm$ 338.94 & 1.50 $\pm$ 2.01 & --6.93 $\pm$ 2.73 \\ 
HAT-P-11\,b & 7.62 & 6.61 & 0.43 $\pm$ 0.01 & 632.37 $\pm$ 228.12 & 4.04 $\pm$ 1.11 & --1.76 $\pm$ 1.41 \\ 
HAT-P-12\,b & 16.12 & 3.08 & 0.92 $\pm$ 0.02 & 509.25 $\pm$ 174.42 & 2.76 $\pm$ 1.23 & --3.61 $\pm$ 1.48 \\ 
HAT-P-17\,b & 5.34 & 0.28 & 0.99 $\pm$ 0.02 & 568.69 $\pm$ 330.38 & 1.25 $\pm$ 2.12 & --5.86 $\pm$ 2.89 \\ 
HAT-P-18\,b & 13.91 & 5.71 & 0.94 $\pm$ 0.02 & 451.61 $\pm$ 176.54 & 2.82 $\pm$ 0.91 & --2.63 $\pm$ 1.18 \\ 
HAT-P-26\,b & 13.59 & 32.73 & 0.52 $\pm$ 0.01 & 680.56 $\pm$ 198.55 & 3.94 $\pm$ 0.74 & --3.32 $\pm$ 1.10 \\ 
HAT-P-32\,b & 14.32 & 16.44 & 1.77 $\pm$ 0.02 & 1139.53 $\pm$ 169.81 & 2.34 $\pm$ 0.88 & --2.84 $\pm$ 0.92 \\ 
HAT-P-38\,b & 5.47 & 0.67 & 0.82 $\pm$ 0.02 & 762.80 $\pm$ 256.24 & 3.32 $\pm$ 1.68 & --4.29 $\pm$ 2.16 \\ 
HAT-P-41\,b & 8.59 & 7.29 & 1.60 $\pm$ 0.03 & 1570.37 $\pm$ 313.42 & 2.41 $\pm$ 1.20 & --2.77 $\pm$ 1.09 \\ 
HD\,149026\,b & 5.82 & 0.00 & 0.65 $\pm$ 0.01 & 1335.30 $\pm$ 379.48 & 0.75 $\pm$ 1.68 & --5.75 $\pm$ 2.91 \\ 
HD\,189733\,b & 7.87 & 11.77 & 1.16 $\pm$ 0.00 & 621.49 $\pm$ 139.05 & 4.66 $\pm$ 0.91 & --2.51 $\pm$ 0.90 \\ 
HD\,209458\,b & 22.24 & 17.21 & 1.33 $\pm$ 0.02 & 1061.35 $\pm$ 241.23 & 2.14 $\pm$ 0.95 & --3.19 $\pm$ 0.87 \\ 
WASP-12\,b & 14.72 & 25.08 & 1.86 $\pm$ 0.02 & 1864.01 $\pm$ 202.82 & 2.38 $\pm$ 0.95 & --3.12 $\pm$ 0.92 \\ 
WASP-29\,b & 9.25 & 1.25 & 0.76 $\pm$ 0.02 & 713.48 $\pm$ 311.15 & 3.29 $\pm$ 2.29 & --7.93 $\pm$ 2.38 \\ 
WASP-31\,b & 9.33 & 1.31 & 1.47 $\pm$ 0.03 & 1088.35 $\pm$ 220.16 & 1.79 $\pm$ 1.27 & --3.84 $\pm$ 1.90 \\ 
WASP-39\,b & 22.66 & 34.52 & 1.24 $\pm$ 0.01 & 1258.71 $\pm$ 389.53 & 4.86 $\pm$ 0.32 & --5.94 $\pm$ 0.61 \\ 
WASP-43\,b & 7.34 & 1.93 & 0.94 $\pm$ 0.01 & 957.27 $\pm$ 343.30 & 2.90 $\pm$ 2.12 & --4.36 $\pm$ 2.10 \\ 
WASP-52\,b & 13.74 & 20.32 & 1.27 $\pm$ 0.01 & 667.66 $\pm$ 121.94 & 4.84 $\pm$ 0.88 & --4.09 $\pm$ 0.87 \\ 
WASP-63\,b & 12.22 & 0.00 & 1.36 $\pm$ 0.03 & 948.22 $\pm$ 179.13 & 0.93 $\pm$ 1.40 & --5.81 $\pm$ 2.81 \\ 
WASP-67\,b & 5.87 & 0.27 & 1.36 $\pm$ 0.03 & 636.58 $\pm$ 267.82 & 2.18 $\pm$ 1.91 & --6.17 $\pm$ 2.82 \\ 
WASP-69\,b & 31.39 & 13.30 & 1.01 $\pm$ 0.01 & 492.92 $\pm$ 153.38 & 3.93 $\pm$ 0.99 & --3.94 $\pm$ 1.25 \\ 
WASP-74\,b & 8.35 & 0.00 & 1.46 $\pm$ 0.03 & 1519.36 $\pm$ 310.70 & --0.05 $\pm$ 1.48 & --5.91 $\pm$ 2.81 \\ 
WASP-76\,b & 23.24 & 36.44 & 1.68 $\pm$ 0.02 & 1591.88 $\pm$ 184.08 & 3.93 $\pm$ 1.22 & --2.70 $\pm$ 1.07 \\ 
WASP-80\,b & 15.75 & 1.16 & 0.98 $\pm$ 0.01 & 539.39 $\pm$ 278.81 & 2.17 $\pm$ 1.48 & --5.34 $\pm$ 2.65 \\ 
WASP-101\,b & 14.03 & 0.00 & 1.29 $\pm$ 0.02 & 1042.55 $\pm$ 215.30 & 0.54 $\pm$ 1.75 & --6.95 $\pm$ 2.61 \\ 
WASP-121\,b & 15.96 & 11.52 & 1.69 $\pm$ 0.01 & 1543.93 $\pm$ 134.06 & 3.79 $\pm$ 1.25 & --3.05 $\pm$ 0.87 \\ 
XO-1\,b & 4.97 & 3.15 & 1.21 $\pm$ 0.01 & 778.21 $\pm$ 224.04 & 4.14 $\pm$ 1.29 & --2.75 $\pm$ 1.64 \\ 
	\end{tabular}
\end{table*}

\begin{figure}
	\centering
	\includegraphics[width=0.96\columnwidth]{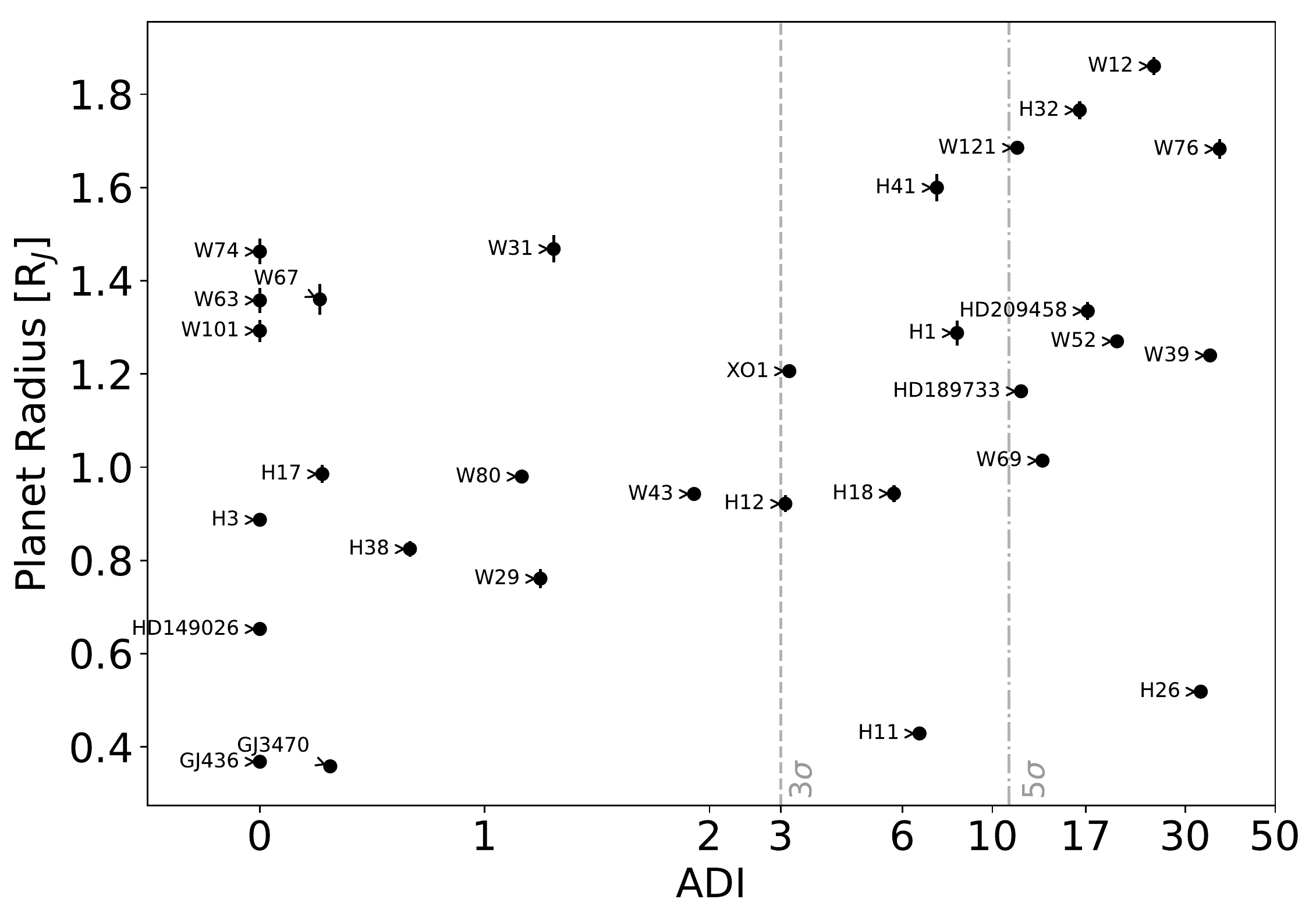}
	\caption{A positive correlation exists between the planet radius and ADI, with larger planets generally featuring more detectable atmospheres. However, We note an outlying cluster of five planets, including WASP-31\,b, WASP-63\,b, WASP-67\,b, WASP-74\,b and WASP-101\,b. These low ADIs may indicate high-altitude cloud covers, or water depleted atmospheres.}
	\label{fig:adi_radius}
\end{figure}

\begin{figure}
	\centering
	\includegraphics[width=0.96\columnwidth]{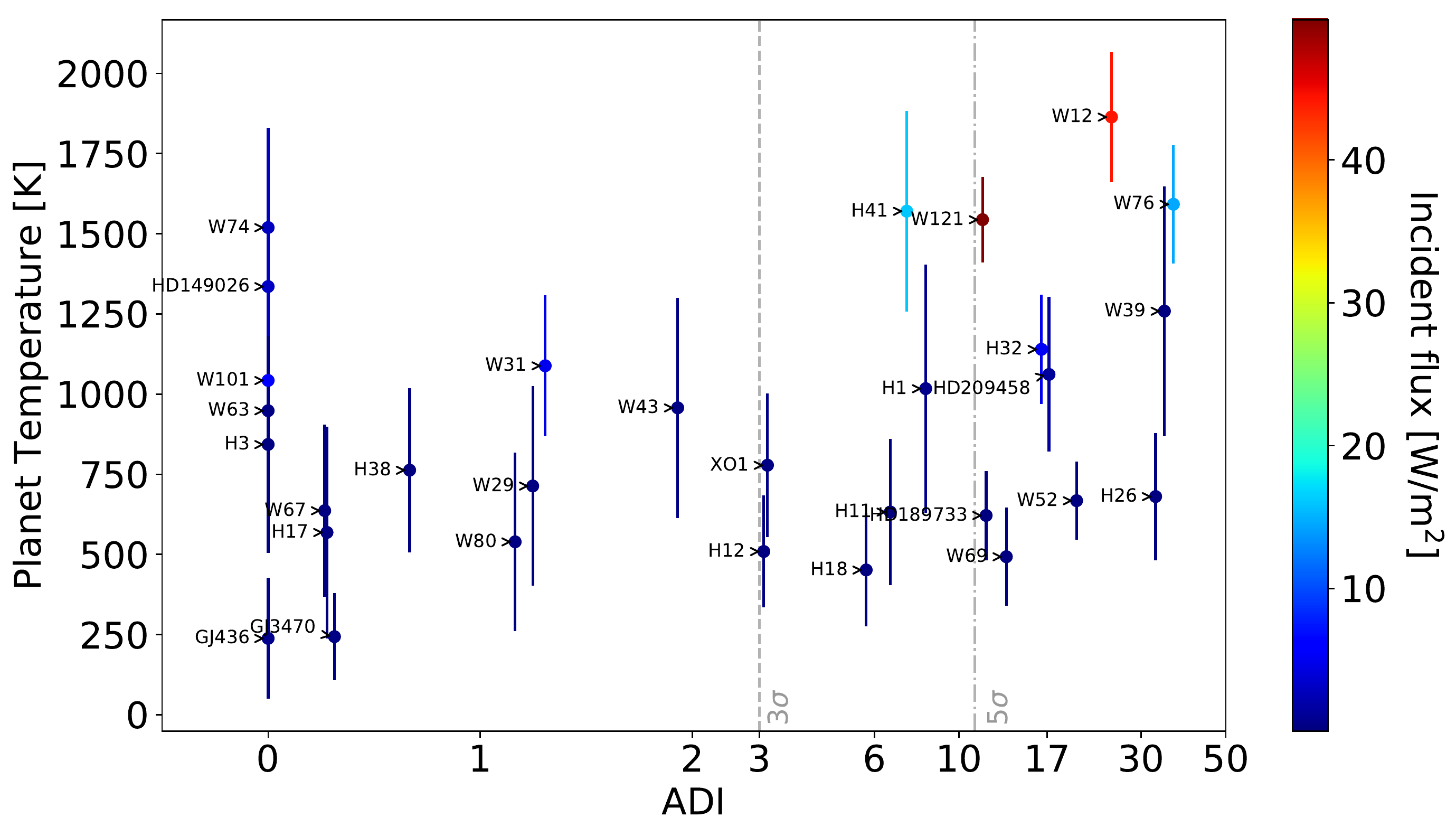}
	\caption{Correlation between retrieved planet temperature and ADI. Colours show the UV radiation the planet receives in W/m$^2$. A cluster of outliers at high temperature and high ADI is apparent. These planets are also the highest irradiated.}
	\label{fig:adi_temp}
\end{figure}

\begin{figure}
	\centering
	\includegraphics[width=0.96\columnwidth]{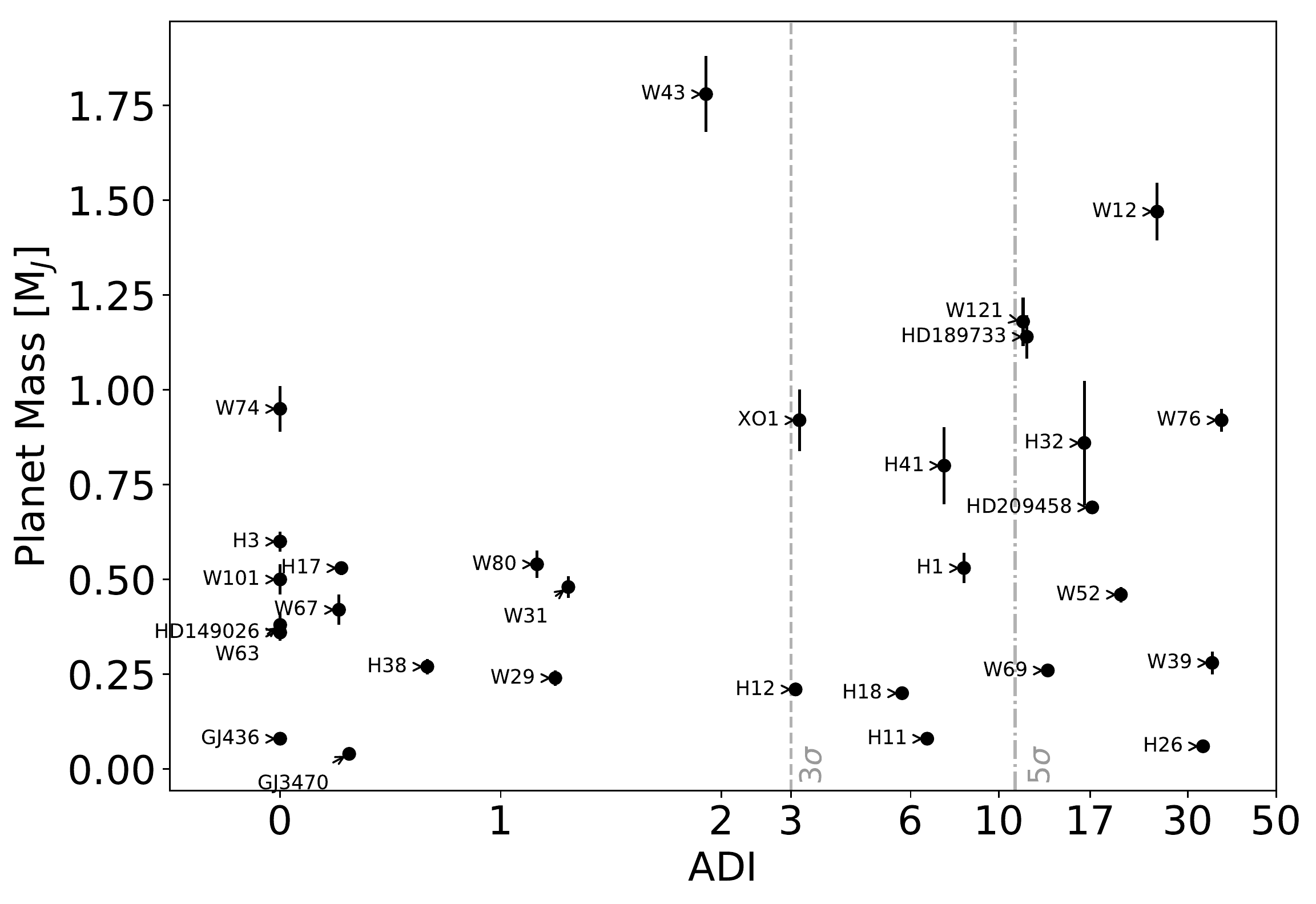}
	\caption{Planetary mass as a function of ADI. While the two groups of planets are clearly separated (with or without detectable atmospheres) there is no evident correlation between the planetary mass and the ADI index.}
	\label{fig:adi_mass}
\end{figure}

\begin{figure}
	\centering
	\includegraphics[width=0.96\columnwidth]{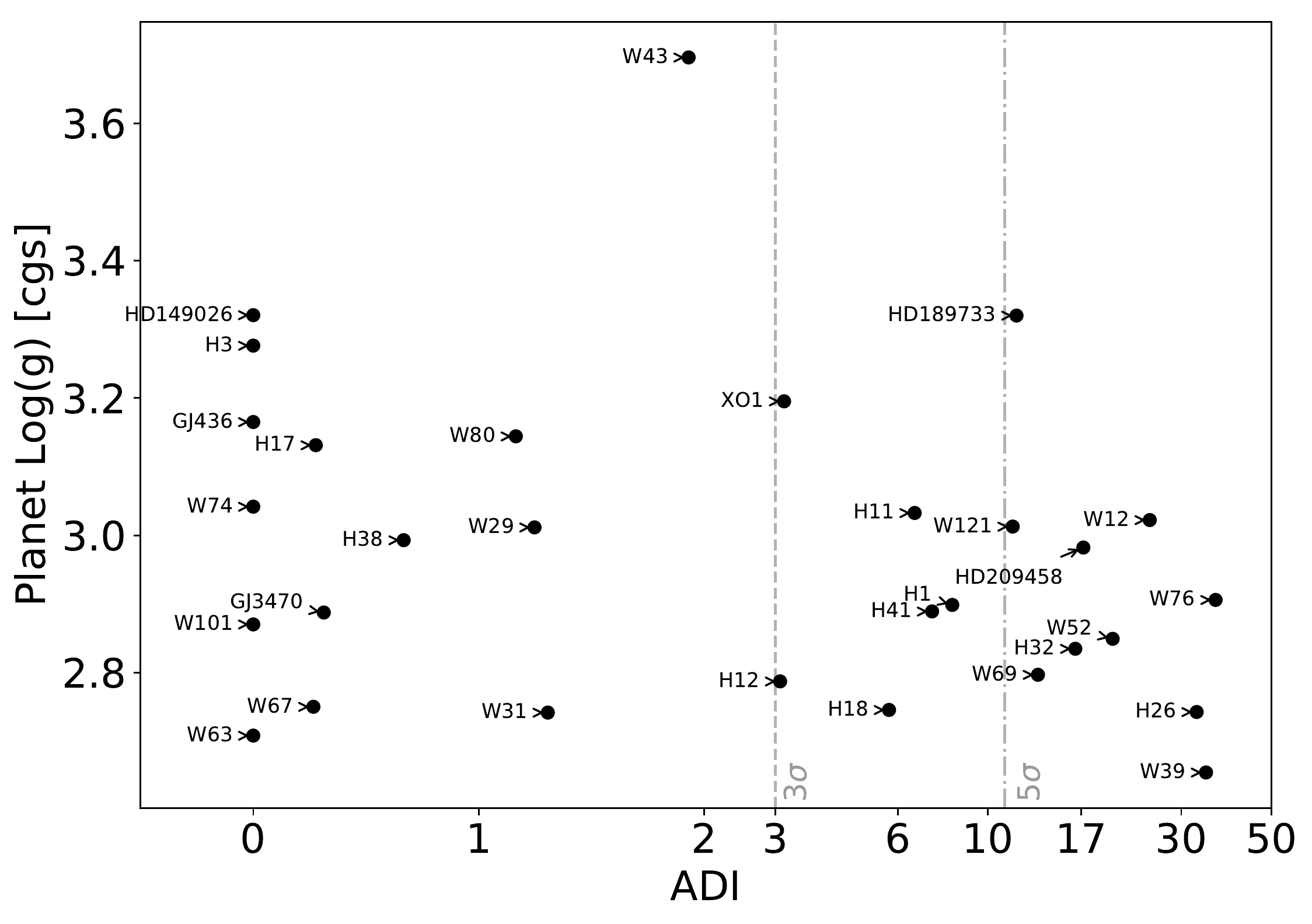}
	\caption{Planetary gravity as a function of ADI, with a similar behaviour to the planetary mass.}
	\label{fig:adi_logg}
\end{figure}

Considering the warm and hot Jupiters in our sample ($M > 0.16\,M_\mathrm{Jup}$, i.e. excluding the Neptunes: GJ\,436\b, GJ\,3470\,b, HAT-P-11\,b and HAT-P-26\,b), the Pearson correlation coefficient indicates that the ADI is more strongly correlated with the planetary radius (0.51, p-value=0.7\%) than the planetary temperature (0.43, p-value=3\%) but not correlated with the surface gravity (-0.28, p-value=16\%) or the planetary mass (0.20, p-value=32\%). These parameters are plotted against ADI in Figures \ref{fig:adi_radius}, \ref{fig:adi_temp}, \ref{fig:adi_mass}, and \ref{fig:adi_logg}. These results indicate that planetary surface gravity is a secondary factor in identifying inflated atmospheres \citep{Laughlin2011, Weiss2013, Spiegel2013}. 

Very hot and highly irradiated planets, with atmospheric temperatures above 1800\,K feature high ADI atmospheres. Our quantitative retrievals suggest that the cloud top-pressures in these planets are significantly high, meaning clouds are deep in the atmosphere, if present at all (Table \ref{tab:all_results}), while retrieved water abundances are constant within the errors. Whilst we cannot determine the absolute atmospheric water abundances, given the relative narrow wavelength range probed, we can exclude scenarios where water is significantly destroyed or depleted in the upper atmospheres of irradiated and inflated hot-Jupiters. In addition, the spectra of HAT-P-41\,b, WASP-12\,b and WASP-121\,b show no contribution from photochemical hazes \citep{Zahnle2009, Kopparapu2012, MillerRicciKempton2012}. We can conclude that planets with temperatures higher than 1800\,K feature clear atmospheres in the terminator regions at HST/WFC3 wavelengths.

\begin{figure*}
	\centering
	\includegraphics[width=0.44\textwidth]{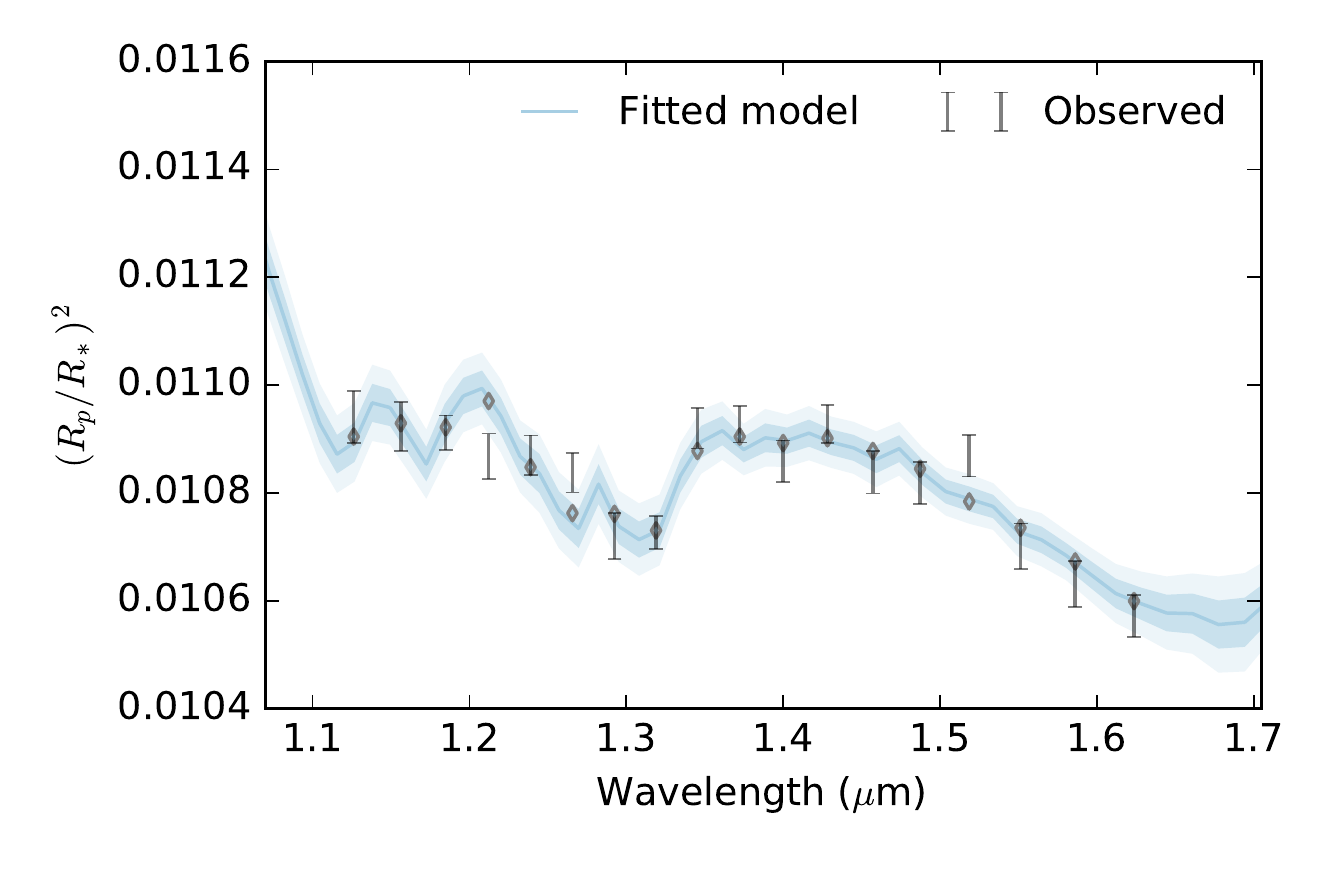}
    \includegraphics[width=0.54\textwidth]{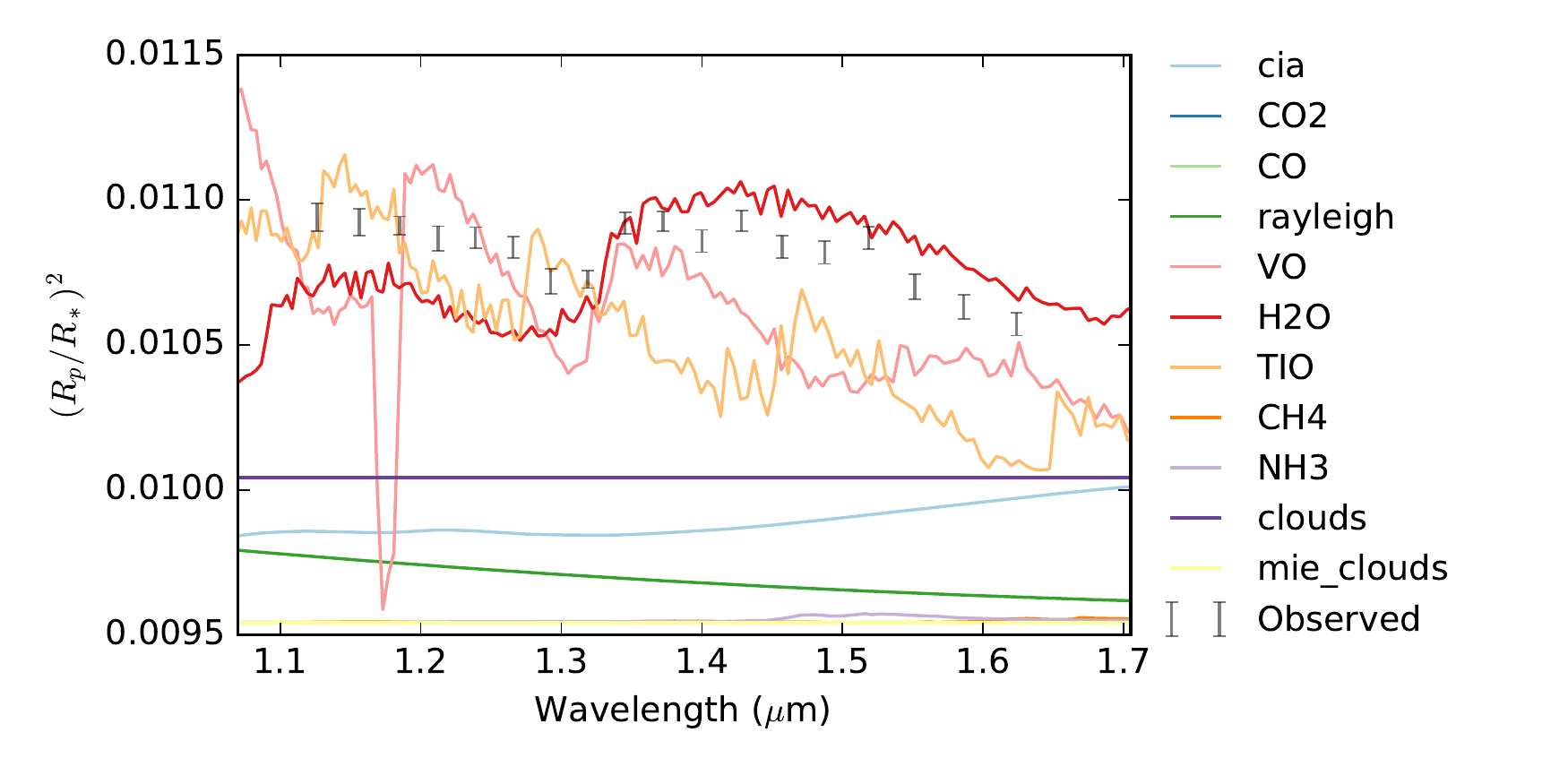}
	\caption{Left: Best fit spectra for WASP-76\,b transmission spectrum in low resolution. A clear (no haze) upper atmosphere with a deep cloud-top (~0.8 bar). Here the main opacities constitute H$_2$O, TiO and VO.}
	\label{fig:wasp76_1}
\end{figure*}

\begin{figure*}
	\centering
	\includegraphics[width=0.7\textwidth]{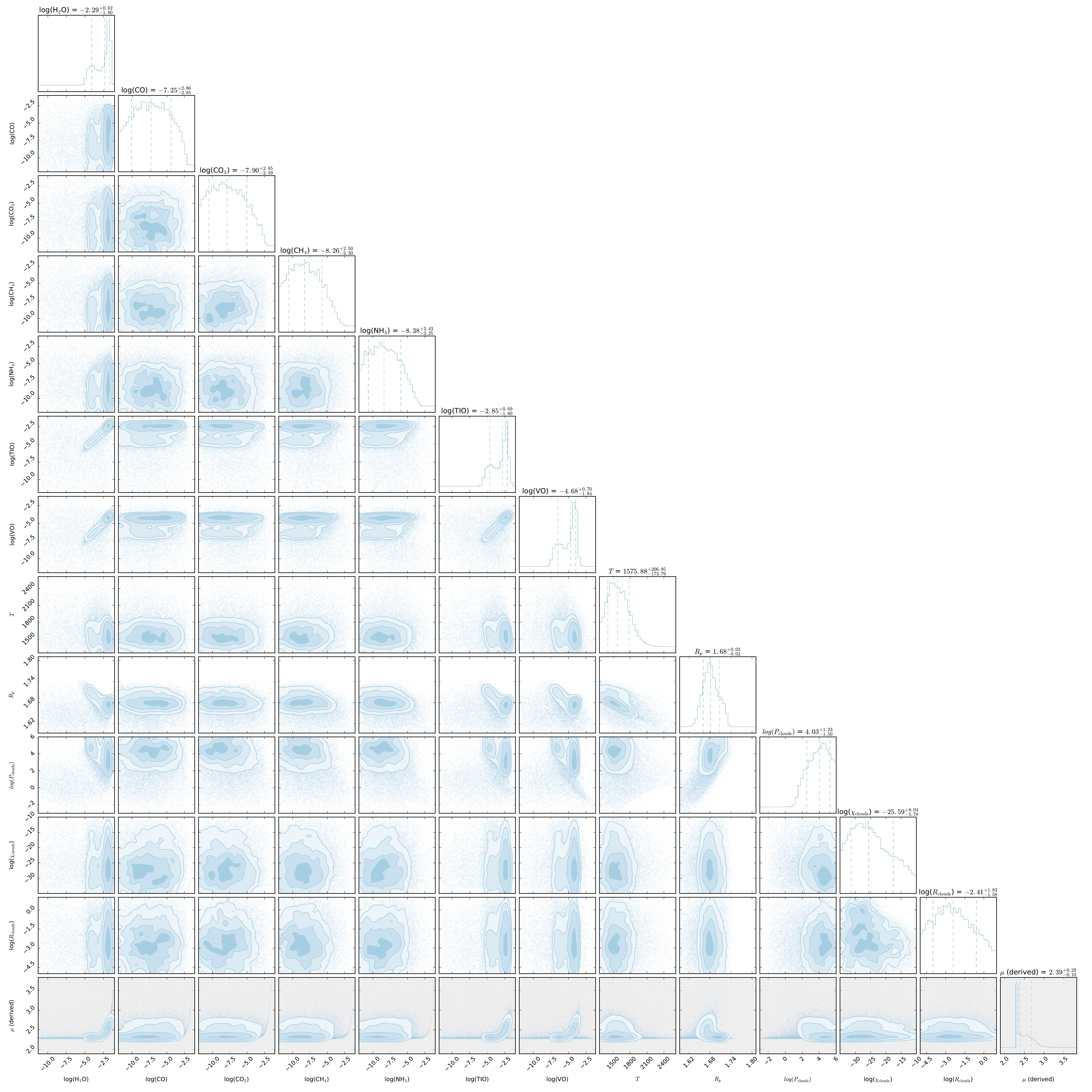}
	\caption{The posterior distribution of the Bayesian retrieval for WASP-76\,b.}
	\label{fig:wasp76_2}
\end{figure*}

In our retrievals we considered a mixture of opaque cloud-deck and hazes, all planets but WASP-69\,b are consistent with a grey, opaque cloud-deck. In this study, both opaque clouds and hazes were uniformly distributed along the terminator. \cite{Line20162016ApJ...820...78L} showed that nonuniform cloud coverage can mimic high-molecular weight (hmw) atmospheres. Whilst hmw atmospheres are not observed in our hot-Jupiter retrievals, we note that \hst/\wfc\ data alone is not sufficient to differentiate between hmw, and low-molecular weight atmospheres with patchy cloud coverage \citep{Line20162016ApJ...820...78L}. This is particularly relevant for the warm-Neptune HAT-P-11\,b, where a hmw atmosphere was postulated by \citet{Fraine2014}. Asymmetric cloud coverage can be observed in ingress/egress signatures of the light-curves \citep{vonParis2016, Line20162016ApJ...820...78L} but the incomplete phase coverage of the HST/WFC3 data is insufficient to confirm or reject patchy cloud coverage models.

\subsection{Molecular opacities detected} 

\begin{figure*}
	\centering
	\includegraphics[width=\textwidth]{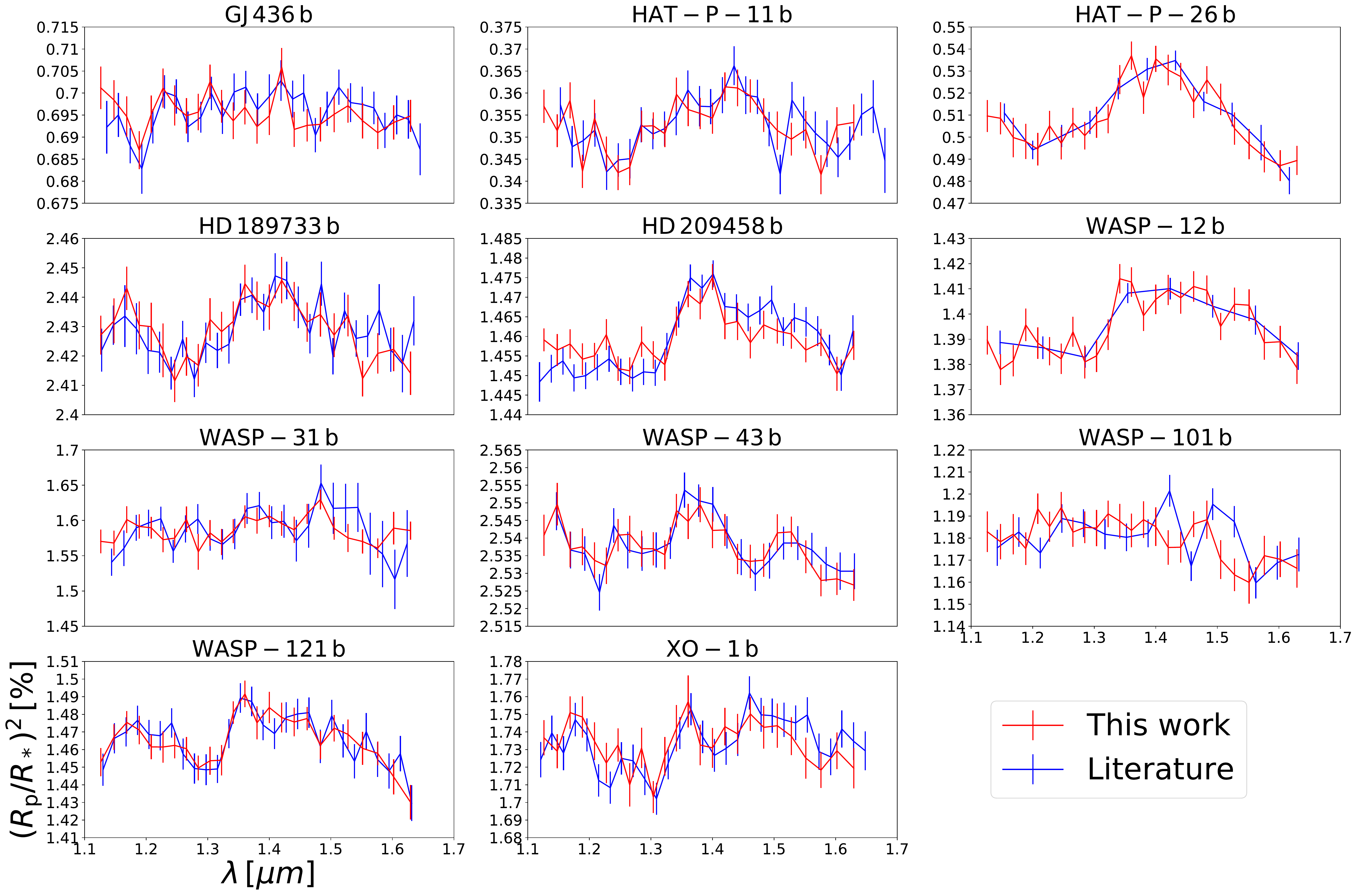}
	\caption{Comparison between the spectra presented here (red) and those available in the literature (blue) for 11 planets in our sample. The spectra have been normalized to have the same average transit depth, as they are subject to arbitrary offsets due to different orbital parameters or limb-darkening coefficients used by different studies.}
	\label{fig:comp_literature}
\end{figure*}

The 16 spectra which show statistically significant atmospheres presented here are well described with a combination of grey-clouds, extended, particulate Rayleigh curves and water. Two notable exceptions are WASP-76\,b (see Figures \ref{fig:wasp76_1} and \ref{fig:wasp76_2}) and WASP-121\,b. Both planets are hot Jupiters with equilibrium temperatures of  $\sim$2000\,K. The retrieval results show that the atmosphere is haze free (i.e. clear) and TiO, VO and H$_2$O opacities determine the observed spectral shape. The TiO \& VO model is favoured with a Bayes Factor of 8.52 (4.44\,$\sigma$ significance) when compared to a pure-water and haze dominated atmosphere for WASP-76\,b. However, we would like to caution the reader that correlations between H$_2$O, TiO and VO abundance, planet radius and cloud-top pressure exist in the retrieved posterior distributions. The retrieval features a high-H$_2$O ($\sim 10^{-2.0}$) and high-TiO ($\sim 10^{-2.5}$) mode, which is likely unphysical. More observations, in particular in the optical wavelengths, are required to fully distinguish between a TiO/VO abundant and high-altitude haze model. In the case of WASP-121\,b, we find both models to be statistically indistinguishable from each other. As discussed above, in this analysis we do not take into account effects due to patchy or non-uniform cloud covers \citep[e.g.][]{MacDonald2017}. In particular, \citet{Kempton2017} shows that non-uniform clouds/hazes on WASP-121\,b can cause observable spectral gradients in the HST/WFC3 wavelengths.

The sparse sampling of \hst/\wfc\ data and the short wavelength ranges do not allow us to conclusively exclude atmospheric haze models for these planets at this stage, though we note that the particulate extended Rayleigh curve would be unusually strong. Observations at longer wavelength ranges are required to conclusively determine the absolute abundance of molecular tracers.

The remaining 14 spectra without a statistically significant atmosphere can be explained by either opaque, high-altitude, clouds or low water abundances, as no-atmosphere models are unlikely for gas-giant planets. Given the uncertainties in the observed spectra, we are sensitive to water mixing ratio higher than 10$^{-8}$, for cloud-free atmospheres. We also note that combinations of water depletion and high-altitude clouds cannot be ruled out. Current space and ground-based data cannot constrain absolute abundances of trace gases beyond their detection. Future instrumentation such as the JWST or dedicated space missions probing a broader wavelength range will be able to break these degeneracies. 

The spectra of 12 out of the 30 planets in our sample have been previously studied. These planets are: GJ\,436\,b \citep{Knutson20142014Natur.505...66K}, HAT-P-1\,b \citep{Wakeford2013}, HAT-P-11\,b \citep{Fraine2014}, HAT-P-32\,b \citep{Damiano2017}, HD\,209458\,b \citep{Deming2013}, HD\,189733\,b \citep{McCullough20142014ApJ...791...55M}, WASP-12\,b \citep{Kreidberg2015}, WASP-31\,b \citep{Sing2015}, WASP-43\,b \citep{Kreidberg20142014ApJ...793L..27K}, WASP-101\,b \citep{Wakeford20172017ApJ...835L..12W}, WASP-121\,b \citep{Evans2016} and XO-1\,b \citep{Deming2013}. Figure \ref{fig:comp_literature} shows a comparison between the extracted spectra here and in the literature. The only noticeable difference is HD\,209458\,b, which we believe is due to the different calibration method used \citep{Tsiaras20162016ApJ...832..202T}. We plan to further investigate this behavior is a future study. Concerning the detection of water vapour and other molecules (TiO, VO) and clouds, our results are consistent with previous results in the literature.

\section{CONCLUSIONS} \label{sec:discussion}

We have presented here the largest catalog of exoplanet atmospheres and atmospheric retrievals to date. Using the most precise data available, analyzed by our specialized tool for WFC3 spatially scanned observations, combined with our fully Bayesian spectral retrieval code and the most accurate molecular line lists, we are able to provide the first fully self-consistent, stable and statistically evaluated reference catalog for comparative exoplanetary characterization.

All software used to create this catalog, and all the intermediate and final data products are publicly available to the community, allowing for reproducibility of the results and further analysis. For more details, visit the UCL Extrasolar Planets page (\href{https://www.ucl.ac.uk/exoplanets}{https://www.ucl.ac.uk/exoplanets}). Also, visit \href{http://bit.ly/HSTDATA}{http://bit.ly/HSTDATA} and \href{https://github.com/ucl-exoplanets}{https://github.com/ucl-exoplanets} to access the datasets and the analysis tools, respectively.

We defined a new metric to estimate the significance of an atmospheric observation, the Atmospheric Detectability Index (ADI). The ADI is the positively-defined logarithmic Bayes Factor between the best-fit water-only model and a grey-cloud/no-atmosphere family of models. It is markedly different to a more classical straight-line rejection as it compares detectable atmospheric features to the full range of possible nondetection models given the data. Amongst the wide diversity of planets, we find about half to have strongly detectable atmospheres featuring water signatures ($\mathrm{ADI} > 3$). We cannot rule out the existence of clouds or water depletion in the remaining, not statistically significant, atmospheres ($\mathrm{ADI} < 3$). Warm and hot Jupiters, with the exception of a distinct group of five hot Jupiters that likely feature very high altitude clouds, follow a clear trend between the ADI and the planetary radius. We find that simple S/N predictions are insufficient for target selection requiring comprehensive spectroscopic observations of targets prior to more detailed studies using large scale observation programs. Population studies such as this one are fundamental in understanding the complex nature and evolutionary history of planets.

\begin{acknowledgements}

This project has received funding from the European Research Council (ERC) under the European Union's Horizon 2020 research and innovation programme (grant agreement No 758892, ExoAI) and under the European Union's Seventh Framework Programme (FP7/2007-2013)/ ERC grant agreement numbers 617119 (ExoLights) and 267219 (ExoMol). We furthermore acknowledge funding by the Science and Technology Funding Council (STFC) grants: ST/K502406/1 and ST/P000282/1.

\end{acknowledgements}

{\small
\bibliographystyle{apj}
\bibliography{references,references_sup} 
}


\newpage


\newpage

\newpage

\newpage


\end{document}